\documentclass[a4paper, amsfonts, amssymb, amsmath, reprint, showkeys, nofootinbib, twoside]{revtex4-2}
\usepackage[english]{babel}
\usepackage{array}
\usepackage[utf8]{inputenc}
\usepackage[colorinlistoftodos, color=green!40, prependcaption]{todonotes}
\usepackage{amsthm}
\usepackage{mathtools}
\usepackage{physics}
\usepackage{xcolor}
\usepackage{graphicx}
\usepackage[left=23mm,right=13mm,top=35mm,columnsep=15pt]{geometry} 
\usepackage{adjustbox}
\usepackage{placeins}
\usepackage[T1]{fontenc}
\usepackage{lipsum}
\usepackage{csquotes}
\DeclareMathSymbol{\mhyphen}{\mathord}{AMSa}{"39}
\usepackage{float}
\restylefloat{table}

\usepackage[pdftex, pdftitle={Article}, pdfauthor={Author}]{hyperref} % For hyperlinks in the PDF
\newcommand{\pvec}[1]{\vec{#1}\mkern2mu\vphantom{#1}}
\begin{document}
\title{
Calculation 
of the
local environment of 
a 
barium monofluoride 
molecule in a
neon matrix}

\author{
R. L. Lambo, 
G. K. Koyanagi, 
M. Horbatsch, 
R. Fournier, 
E. A. Hessels}
\email[Correspondence email address: ]{hessels@yorku.ca}% Your name
\affiliation{
York University,
Toronto ON,
Canada}
\collaboration{
EDM$^3$
collaboration}
\date{\today} % 

\begin{abstract}
The local environment 
of a
barium monofluoride
(BaF)
molecule
embedded in a neon
matrix 
is studied theoretically.
The energy of the 
BaF-Ne
triatomic
system  
is calculated 
with a scalar relativistic 
Hamiltonian,
using
coupled-cluster
theory
at the
CCSD(T) level
for 
$1625$ 
positions
of the
Ne
atom
relative to the
BaF 
molecule.
The calculations 
are repeated with
increasing basis
sets 
(from double
to 
quintuple zeta),
and are 
extrapolated to 
estimate the 
complete-basis-set
limit.
Using the 
potential obtained from 
these calculations, 
it is determined that
substituting 
a 
BaF
molecule for 
ten 
Ne 
atoms is 
favoured
compared to substitutions
for other numbers of 
Ne
atoms.
The 
equilibrium position 
and orientation 
of the 
BaF 
molecule
and
the displacement of
its nearby 
Ne
neighbours
are 
determined.
The potential barriers 
that prevent
the 
BaF 
molecule 
from migrating and rotating
are calculated.
These barriers are
essential for 
the 
EDM$^3$ 
collaboration,
which is using 
BaF 
molecules 
embedded in 
a 
noble-gas
solid to perform
a precision 
measurement of
the electron
electric dipole
moment.

\end{abstract}


\maketitle

\section{
Introduction}

Barium 
monofluoride
embedded in a 
rare-gas
matrix
is being investigated
\cite{vutha2018orientation}
by the 
EDM$^3$
collaboration to
carry out a
high-precision
measurement
of the 
electron electric 
dipole moment
(eEDM).
Current measurements
\cite{roussy2022new, acme2018improved}
of the 
eEDM
already 
test physics at 
energy scales of up to
100~TeV,
and 
put strong limits on possible
beyond-the-standard-model
physics
that would allow for the 
level of
time-reversal (T) violation 
needed to expain the 
asymmetry between matter and
antimatter in the universe.
More precise measurements of the 
eEDM 
will probe for new physics at 
even higher
energies.

As described in 
Section~\ref{sec:triatomic},
we 
use
all-electron
scalar-relativistic
quantum-mechanical
calculations
at the level of
coupled-cluster
theory using the 
second-order
Douglas-Kroll-Hess
method
to 
obtain 
ground-state
energies of the 
BaF-Ne 
triatomic
system.
We 
include 
correlation
at the 
CCSD(T)
level
and  
extrapolate to 
the 
complete-basis-set
limit.
The energies are
calculated for a large
range of positions 
(distances
$r$
and
angles
$\theta$)
of the 
Ne
atom
relative to the
BaF
molecule.
Interpolations provide
a smoothly varying 
potential energy 
versus
$r$
and
$\theta$
for the 
interaction between
BaF
and 
Ne.

As described in 
Section~\ref{sec:environment},
this potential,
along with 
the 
well-known
Ne-Ne
interatomic 
potential
is used to
calculate the 
local environment 
near the embedded
BaF
molecule.
A substitution of 
a 
BaF
molecule for 
ten
Ne
atoms 
is found to be 
favoured
energetically 
compared to substitutions
for other numbers of 
Ne
atoms,
and a second local
minimum is found for 
a substitution of 
thirteen 
Ne
atoms.
The equilibrium 
positions of the
BaF
molecule 
and of the neighbouring
Ne
atoms 
are determined.
Large potential barriers
are found to prevent the 
BaF 
molecule
from migrating through
the solid 
and from 
rotating within
the solid.
These barriers will
allow for 
non-rotating,
non-migrating
BaF
molecules
to be held by the solid,
as is required for 
an
eEDM
measurement
\cite{vutha2018orientation}.

Recently,
BaF
molecules
have been embedded in both
Ne
\cite{li2022baf}
and
Ar
\cite{corriveau2023baf}
cryogenic solids
by the 
EDM$^3$
collaboration.
In both cases,
laser-induced
fluorescence 
is observed
at wavelengths
close to those 
seen for 
the free
BaF
molecule.
Electron-spin-resonance
studies
\cite{knight1971hyperfine,li2022baf}
of
matrix-isolated
BaF
molecules
have also 
been performed.
Knowledge of the 
location and orientation of the 
BaF 
molecule
within the 
fcc 
Ne
crystal
is needed  
to 
understand the underlying 
physics involved in this 
spectroscopy and 
help 
guide continuing
work of the 
EDM$^3$
collaboration.

\section{
Triatomic 
${\rm \textbf{BaF-Ne}}$ 
calculations
\label{sec:triatomic}}

\begin{figure}
\includegraphics
[width=0.9\linewidth]{
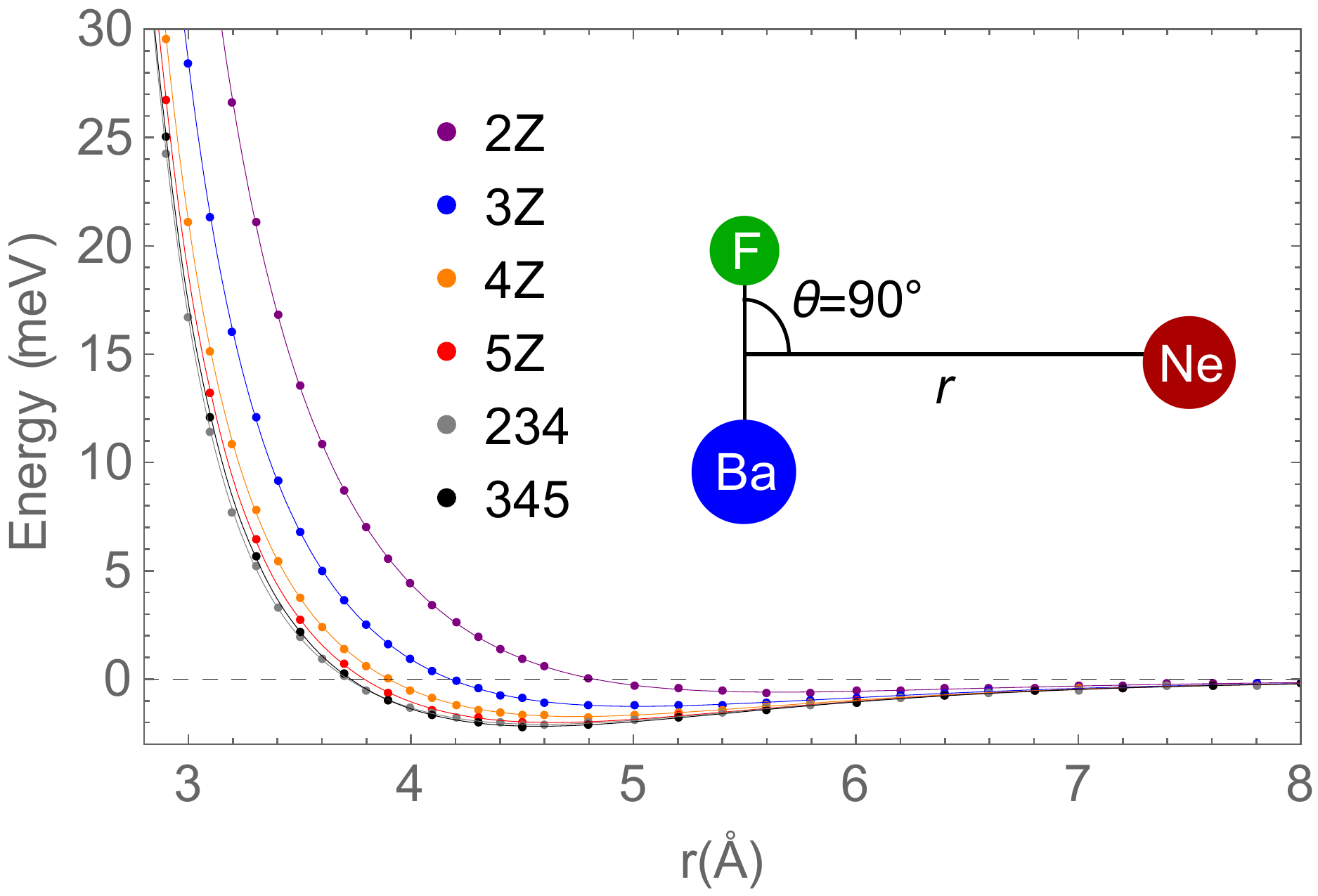}
\caption{
(color online)
The calculated energies of the 
triatomic
BaF-Ne 
system for the case of 
$\theta=90^\circ$.
The curves shown are for 
$\zeta=2$
through
$\zeta=5$
basis sets.
Also shown are the 
complete-basis-set 
extrapolations of 
Eqs.~(\ref{eq:234}) 
and
(\ref{eq:345}).
The level of agreement
between the two extrapolations
leads to an estimate of the
uncertainty in our calculations.
}
\label{fig:extrap90}
\end{figure}

Calculations of 
the 
BaF-Ar
binding energy 
are
performed at 
1625
positions 
of a 
Ne 
atom
relative to a
BaF 
molecule
whose 
internuclear
separation is held fixed
(25 
angles,
$\theta$,
and 
65 
distances,
$2.4$~\AA~$\le r \le16$~\AA)
Here 
$r$
is  
the distance between
the 
Ne 
nucleus
and
the 
midpoint between
the 
Ba
and
F
nuclei
and 
$\theta$
is defined
to be the angle
between the
Ne 
and 
F
nuclei
relative to this 
midpoint, 
as shown in 
the inset in
Fig.~\ref{fig:extrap90}.

The methods 
used are similar to 
those of our previous
work 
\cite{koyanagi2023accurate}
on the 
BaF-Ar 
triatomic
system.
For 
F
and 
Ne,
non-relativistic 
all-electron 
$\zeta$$=$$2$
through
$5$
basis sets 
(aug-cc-pV$\zeta$Z) 
are obtained
from 
Kendall et al.
\cite{kendall1992electron}
and
Woon et al. 
\cite{woon1993gaussian}.
These
basis sets are 
fully 
uncontracted 
and atomic calculations
are performed at the 
Douglas-Kroll-Hess 
(DKH) 
second-order
scalar-relativistic level
\cite{jansen1989revision,wolf2002generalized}. 
Using
orbital occupations
from this 
calculation, 
the basis sets are 
re-contracted 
to their
original structure. 
The basis sets for 
Ba 
are an extension of
the work of 
Hill and Peterson 
\cite{hill2017gaussian},
as
described in 
Ref.~\cite{koyanagi2023accurate}.
All calculations are performed
using the 
Gaussian~16
suite of programs
\cite{frisch2020gaussian}
on 
Digital Research Alliance of Canada
computers.
% and were confirmed with 
% the 
% open-source
% Psi4
% program
% \cite{turney2012psi4}.

The binding energies 
calculated are
$E($BaF-Ne$)-E($BaF$)-E($Ne$)$.
% The use of 
% a finite basis set results
% in 
% $E($BaF-Ne$)$
% being artificially low 
% due to diffuse
% Ne
% orbitals improving 
% the core description
% of
% BaF
% and diffuse 
% Ba
% and 
% F 
% orbitals improving 
% the core description of 
% Ne.
% To correct for this
Counterpoise corrections
are applied
(as described in 
Ref.~\cite{koyanagi2023accurate}), 
whereby 
$E($Ne$)$ 
is computed with
the inclusion of 
the basis functions 
of the 
Ba
and
F
atoms
% BaF 
% ghost orbitals 
and 
$E($BaF$)$
is computed with 
the inclusion of 
Ne
% ghost
% orbitals
basis functions 
\cite{boys1970calculation}.

Correlation is incorporated using 
the 
coupled-cluster 
method 
CCSD(T) 
\cite{raghavachari1989fifth,bartlett1990non,stanton1997ccsd}.
% The 
% correlation-consistent
% basis sets are
% designed to 
% recover 
% the
% correlation energy
% in a consistent
% manner across 
% atom 
% centres
% and in a
% progressive manner 
% as valence and
% polarization
% orbitals are added. 
% As such, 
% extrapolation
% to
The total correlation
energy 
can be obtained
by studying the
problem with a
succession of
basis sets and 
extrapolating to a
complete-basis-set 
(CBS).
By comparing an
extrapolation 
\cite{buchachenko2018interaction}
\begin{eqnarray}
E_{234}
\!=\! 
1.677
E_4
\!-\! 
0.712
E_3
\!+\! 
0.035
E_2
\label{eq:234}
\end{eqnarray}
of the 
energies
using 
$\zeta=2$,
$3$
and
$4$
basis sets
to the analogous extrapolation
from
$\zeta=3$,
$4$
and
$5$,
\begin{eqnarray}
E_{345}
\!=\!
1.593 E_5
\!-\!
0.597 E_4 
\!+\!
0.004 E_3,
\label{eq:345}
\end{eqnarray}
we are able to estimate
the uncertainty of our calculations.
Other methods of 
extrapolating to 
a complete basis set 
are discussed in the 
literature
\cite{feller1992application,peterson1994benchmark,martin1996ab,klopper1999highly}.
As in 
Ref.~\cite{koyanagi2023accurate},
we chose the extrapolations 
of 
Eqs.~(\ref{eq:234})
and
(\ref{eq:345})
since they give good agreement with
measured quantities in  
BaF,
Ba
and
Ba$^+$.
From the level of agreement,
we estimate the uncertainty 
of our
$E_{345}$
calculations to be 
one quarter of the 
$E_{345}-E_{234}$
difference
(as detailed in 
Ref.~\cite{koyanagi2023accurate}).

Figure~\ref{fig:extrap90} 
shows 
$\zeta$$=$$2$
through
$5$
calculations
for the 
BaF-Ne 
system
for the case of 
$\theta=90^\circ$,
along with the two
extrapolations.
Note that the difference between the
two extrapolations is small
($\sim$1~meV  
at
3~\AA,
$\sim$0.1~meV
at 
5~\AA,
and 
$\sim$0.01~meV
at
7~\AA).
The 
computationally-expensive
$\zeta$$=$$5$
calculations were performed
at 
525
carefully-chosen 
points
and,
at points where 
the 
$\zeta$$=$$5$
calculation
were not performed,
$E_{345}$
is 
estimated to high accuracy
from a smooth interpolation
and extrapolation 
of 
$E_{345}-E_{234}$
obtained from the
525 points.
% We take 
% $E_{345}$
% as our best 
% estimate of the 
% potential and 
% use the
% difference 
% between
% $E_{234}$
% and
% $E_{345}$
% to get the scale 
% of the uncertainties
% associated with the 
% CBS
% extrapolations.
% Using 
% comparisons with
% measured quantities in  
% BaF,
% Ba,
% Ba$^+$
% (as detailed in 
% Ref.~\cite{koyanagi2023accurate})
% and 
% Ar,
% as shown in 
% Table~\ref{table:Compare}
% and discussed below,
% we
% estimate 
% the 
% CBS
% extrapolation
% uncertainty 
% to be approximately
% one quarter of 
% the difference 
% between
% $E_{234}$
% and
% $E_{345}$.

The current 
calculations are 
performed with the 
internuclear 
Ba-F
separation
fixed.
The 
BaF
binding energy 
(6 eV
\cite{ehlert1964mass,hildenbrand1968mass})
is large compared to the 
BaF-Ne
binding energy 
(8 meV; 
calculated here),
which leads to a much 
stronger restoring 
force for 
BaF 
stretching
compared to 
BaF-Ne
interactions
and
justifies
fixing 
the
BaF
internuclear 
separation.
The separation is fixed at
2.16~\AA~--
the separation 
determined from 
rotational spectroscopy
\cite{bernard1992laser}.
Extrapolation of our 
CCSD(T) 
diatomic potential
curves is consistent
\cite{koyanagi2023accurate}
with 
this value.

\begin{figure}
\includegraphics
[width=1.0\linewidth]{
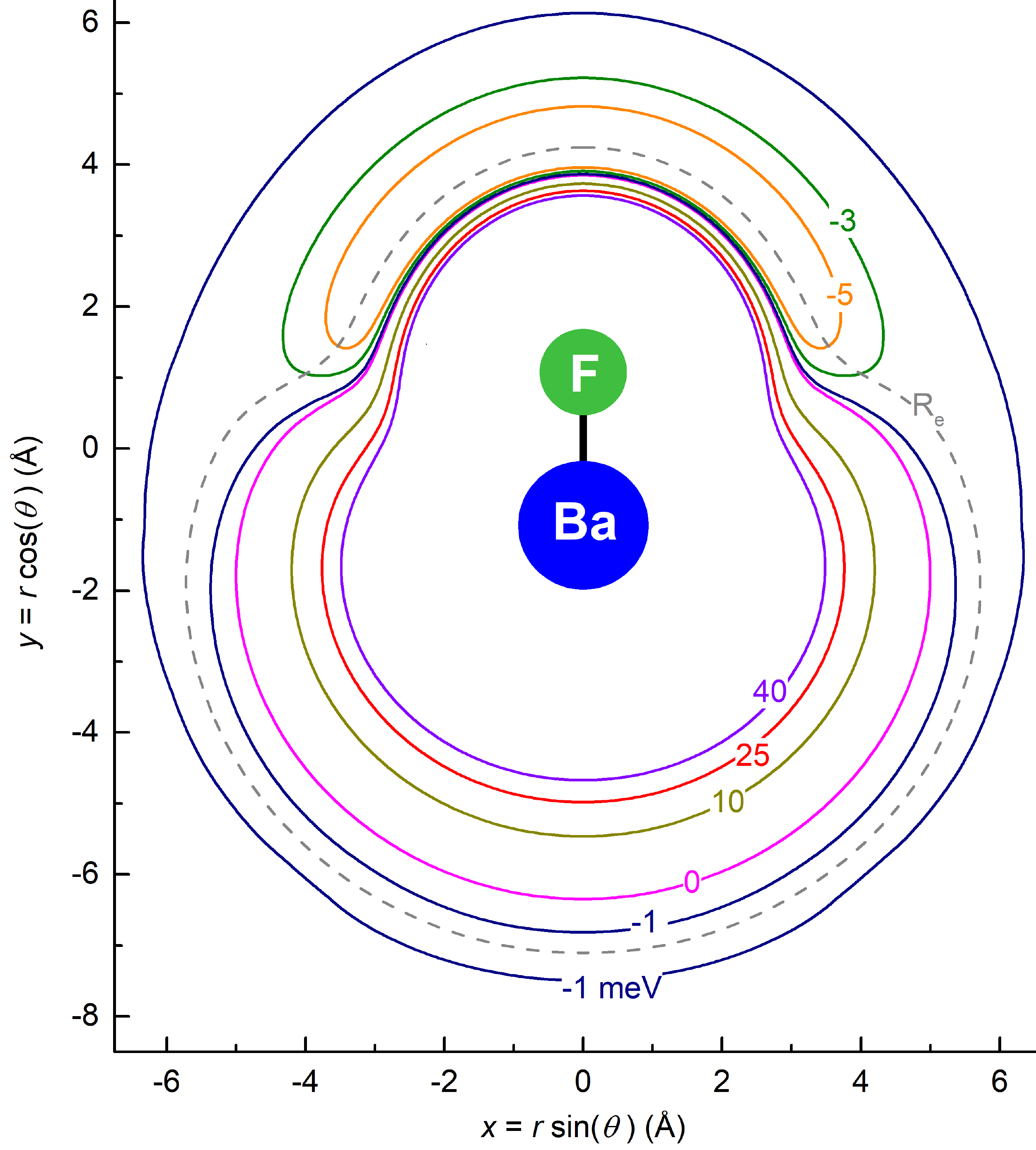}
\caption{
(color online)
A contour graph of 
the calculated 
potential energy between 
a
Ne
atom and a 
BaF 
molecule.
The geometric centre of the 
Ba 
and 
F
nuclei
is situated at 
the origin 
of the plot.
The dashed line
shows the equilibrium
separation 
$R_{\rm e}$
as a function of 
$\theta$.
}
\label{fig:contours}
\end{figure}

\begin{figure*}[hbt!]
    \includegraphics[width=1.0\linewidth]
    {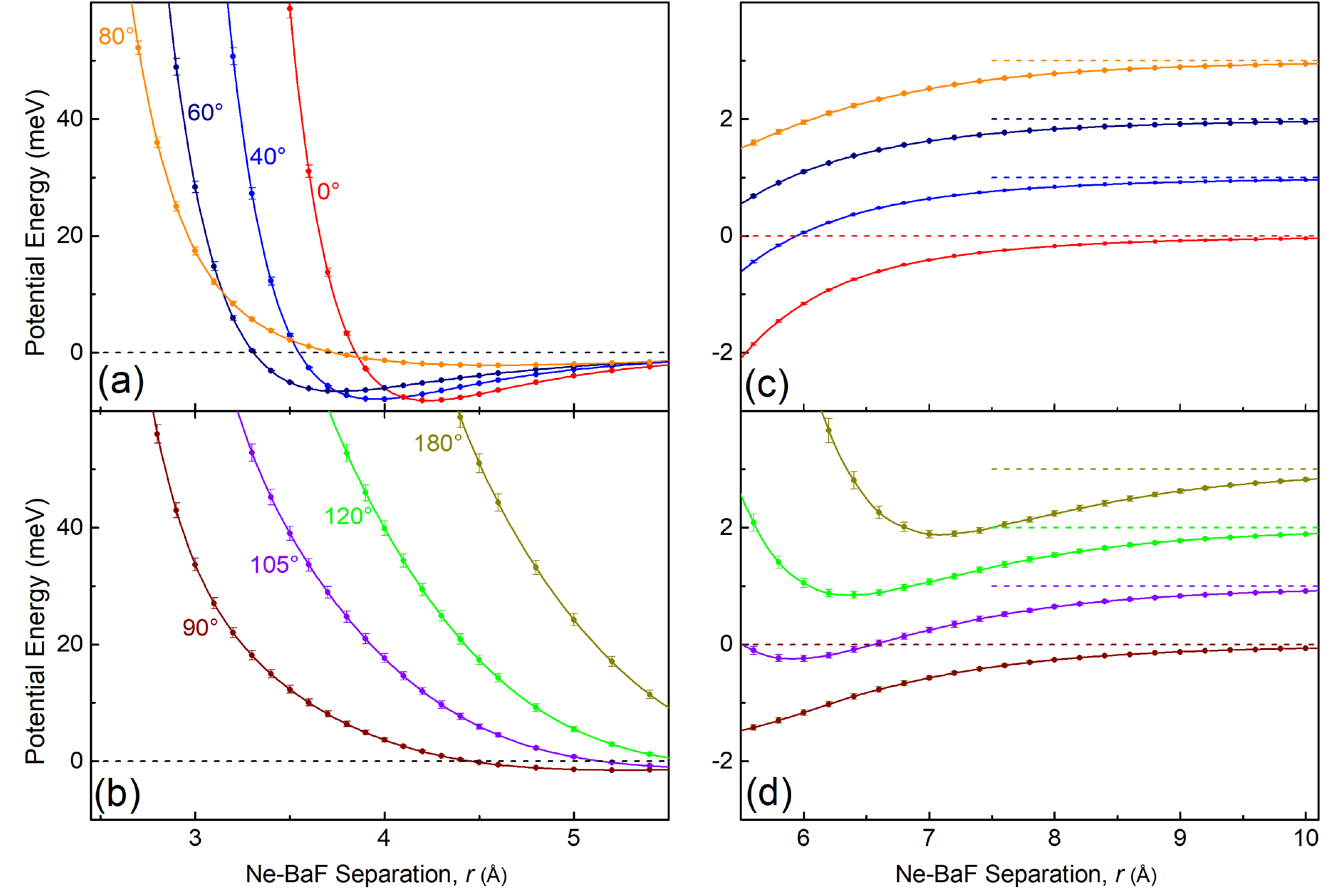}
    \caption{
    (color 
    online)
    The potentials shown for
    8
    of the 
    25
    angles,
    $\theta$,
    calculated here.
    The points plotted 
    are the 
    $E_{345}$
    extrapolations,
    and the uncertainties
    are one quarter of the 
    typical 
    $E_{345}-E_{234}$
    difference at this 
    energy.
    % The 
    % %%%delete "red"
    % fit curves 
    % are 
    % as described in 
    % Section~\ref{sec:fit}.
    The potential curves 
    at longer distances 
    shown in 
    (c)
    and
    (d)
    are offset for clarity.
    }
    \label{fig:potentials}
\end{figure*}

Figure~\ref{fig:contours} 
shows a contour plot of 
our calculated 
BaF-Ne
potential and
Figure~\ref{fig:potentials}
shows the calculated
BaF-Ne
energies for 
8
of the
25
angles
$\theta$
used for
our calculations.
Parts 
(a)
and
(b)
of 
Fig.~\ref{fig:potentials}
show the 
shorter-distance
potentials
and parts
(c)
and 
(d) 
show the 
longer-distance 
potential, 
and these have 
their zeros offset
by 
1,
2
and 
3~meV
for clarity.
The uncertainties shown 
in the figure are based
on one quarter of typical 
$E_{345}-E_{234}$
differences,
with uncertainties of 
approximately
1~meV
at 
25~meV on the repulsive wall,
0.1~meV
at 
$E_{345}=0$,
0.05~meV
at 
the 
potential minimum,
and reducing to 
0.0005~meV 
at long distances.
The full set of 
calculated
energies is 
given in 
Tables~\ref{table:sup0}
through
\ref{table:sup180}
of the Supplementary
Materials.

As can be seen in 
Fig.~\ref{fig:contours},
the 
Ne-BaF
potential varies 
strongly with angle.
For 
$\theta=0$,
for which the 
Ne
atom
is on the 
F
side of the 
BaF
molecule,
the binding energy is 
larger
($D_{\rm e}=$
8.2~meV
at 
$R_{\rm e}\approx$4.2~\AA),
whereas
on the 
Ba
side,
the
Ne 
atom is bound more weakly 
and
the equilibrium 
distance is much larger
($D_{\rm e}=$
1.1~meV 
at
$R_{\rm e}\approx$
7.1~\AA).
The 
short-distance repulsion has
steep curves at
small
$\theta$
and 
increasingly 
gentle curves for
larger 
$\theta$.
The 
large 
Ba
atom leads to 
the repulsive wall
of the potential energy 
being at larger 
distances 
(with a value of
40~meV 
at
approximately 
4.8~\AA)
and
the smaller
F
atom allows for 
closer separations
(40~meV
at 
approximately
3.6~\AA).
The repulsive wall is 
even closer near 
$\theta=90^\circ$,
where 
40~meV occurs
at approximately
2.9~\AA.

\section{
Local environment of 
${\rm \textbf{BaF}}$
in a 
${\rm \textbf{Ne}}$ 
solid
\label{sec:environment}}

Using the 
triatomic
calculations of 
Section~\ref{sec:triatomic},
the geometry and energy of 
a 
BaF
molecule 
in a
Ne
crystal can be determined.
We employ methods 
similar to those 
used in our 
previous work
\cite{lambo2023calculation},
in which the  
BaF-Ar
system is discussed.
Solid 
Ne
has a 
face-centred-cubic 
(fcc)
structure with a cube 
dimension of 
$a=4.4644$~\AA~\cite{batchelder1968isotope}
for 
$^{20}$Ne,
$4.4559$~\AA~\cite{batchelder1968isotope}
for 
$^{22}$Ne,
and 
$4.4637$~\AA~\cite{batchelder1967measurements}
in the
naturally-occurring
mixture of isotopes.
% Relative to one
% Ne
% atom, 
% there are 
% $n_1=12$ 
% nearest neighbours
% at
% $b_1=a/\sqrt{2}$,
% with subsequent 
% sets 
% (of size
% $n_k$)
% of nearest 
% neighbours
% at 
% $b_k=\sqrt{k} b_1$
% (for 
% subsequent
% integer values 
% $k$).
Our simulations use a 
cluster of 
Ne
atoms
within a sphere of
radius 
$b_n=\sqrt{n}a/\sqrt{2}$
(the distance between 
$n^{\rm th}$-nearest
neighbours in the crystal).
Similar to  
Ref.~\cite{tao2015heat},
the outer part of this sphere 
(those farther than  
$b_m$
from the centre)
have their positions 
fixed at the ideal
Ne
fcc 
crystal positions.
Inside this shell,
a single 
BaF
molecule 
is situated near the 
centre,
surrounded by 
$M$ 
Ne
atoms.
The 
$M$
Ne
atoms
and 
one
BaF
molecule 
are allowed to move
to minimize
the overall 
interaction energy of 
the system.

The total energy 
of the cluster is 
minimized while 
varying 
$3M+5$ 
parameters:
the positions of the 
$M$
Ne
atoms
and of the 
BaF molecule,
as well as
the
two angles 
defining
the 
orientation of the 
BaF 
molecule.
As in 
Ref.~\cite{lambo2023calculation},
two independent 
calculations 
using different methods
are employed for this 
minimization to 
verify that the 
global minimum is found.
As  
discussed above,
the 
separation between the 
Ba
and
F
nuclei 
is fixed
at 
2.16~\AA,
the separation determined 
from rotational spectroscopy
\cite{bernard1992laser}.
% The
% large
% BaF 
% binding energy 
% (6 eV 
% \cite{ehlert1964mass,hildenbrand1968mass}) 
% compared to
% the 
% BaF-Ar 
% binding energy 
% (23 meV 
% \cite{koyanagi2023accurate}),
% along with the larger
% equilibrium separations 
% for the 
% BaF-Ar
% system,
% leads to a much stronger 
% restoring force for 
% stretching this 
% 2.16~\AA~separation
% compared to 
% typical 
% BaF-Ar forces,
% justifying 
% a 
% fixed
% BaF 
% internuclear separation.

% As in 
% Ref.~\cite{lambo2023calculation},
% two independent 
% calculations 
% using different methods
% are employed for this 
% minimization to 
% verify that the 
% global minimum is found.
% One hundred
% independent simulated annealing runs
% with different random 
% initial configuration 
% and different cooling schedules
% from a temperature
% $T_{\rm high}$
% (of between 
% 50 
% and 
% 100~K)
% to 
% $T_{\rm low} < 0.1$~K
% are carried out 
% for each energy 
% minimization performed.
% Averaging the 
% low-$T$ 
% configurations
% within a simulation 
% yielded the 
% lowest-energy 
% configuration for that run.
% The five 
% lowest energies found among 
% the 
% 100 
% runs are typically within a few
% meV.
% The lowest of these 
% is further refined 
% by local minimization 
% to an accuracy of
% better than 
% 0.1~meV.
% An independent 
% program for minimization uses 
% $\approx 10^5$ 
% trials 
% with randomly chosen
% initial positions and 
% an adaptive gradient search.
% The lowest twenty energies
% obtained from these trials
% agree to better than
% 1~meV,
% and these results
% (their energies and 
% positions)
% agree with the annealing 
% results.
The value of $M$
is chosen to be 
$S$
fewer than the 
number of atoms that 
would fully occupy
the sphere
of radius
$b_m$,
which allows the 
BaF
molecule to substitute
for 
$S$
Ne 
atoms.
The total number of 
Ne
atoms in the 
simulations is
$N$,
including 
an outer shell
of 
$N-M$
fixed 
atoms extending to a radius 
of 
$b_n$.

The energy being minimized 
is the sum of 
pairwise interactions:
\begin{eqnarray}
E
=&&
\sum_{i=1}^{M}\sum_{j=i+1}^{N}
V_{\rm Ne \mhyphen Ne}(|\Vec{r}_i-\Vec{r}_j|)
\nonumber
\\
+&&
\sum_{i=1}^{N}
V_{\rm BaF \mhyphen Ne}(|\Vec{r}_i-\Vec{r}_0|,\theta_i),
\label{eq:totalE}
\end{eqnarray}
where
$\Vec{r}_0$
is the position of the 
BaF molecule,
$\Vec{r}_i$
is the position of the 
$i^{\rm th}$
Ne 
nucleus,
and 
$\theta_i$
is the angle between the 
internuclear 
axis
($\Vec{r}_{\rm F}-\Vec{r}_0$)
and 
the 
$i^{\rm th}$
Ne 
atom
($\Vec{r}_i-\Vec{r}_0$).

The interaction energy between two 
Ne
atoms,
$
V_{\rm Ne \mhyphen Ne}
(r)
$
is precisely known
\cite{hellmann2008ab}.
To correctly describe 
a 
Ne
crystal, 
however,
corrections
must be included
\cite{rosciszewski2000ab}
to these 
Ne-Ne
interactions.
The dominant correction 
is due to
the
zero-point
energy
of the 
Ne 
atoms,
which requires
an averaging of 
$
V_{\rm Ne \mhyphen Ne}
(r)
$
over the 
positional 
probability 
distributions 
that results from the 
zero-point
motion
of
the atoms.
The next-largest correction 
is due to  
three-body 
Ne-Ne-Ne
interactions.
Without these corrections, 
the calculated 
lattice constant 
and cohesive energy
for the 
Ne
fcc 
crystal would be incorrect.
One approach to compensating 
for these effects follows
the example of 
Ref.~\cite{bezrukov2019empirically},
by using a scaled potential 
\begin{equation}
V^{\rm mod1}_{\rm Ne \mhyphen Ne}(r)
=
\alpha V
_{\rm Ne \mhyphen Ne}
(\beta r).
\label{eq:scaled}
\end{equation}
Coefficients 
$\alpha=0.7048$
and
$\beta=0.9587$
are then chosen 
to match the 
experimental 
Ne
fcc 
cube 
dimension of 
$b=4.4637$~\AA~\cite{batchelder1967measurements}
and 
cohesive energy
of 
$E_{\rm coh}=20.03$~meV
\cite{mcconville1974new}.

However, 
since the corrections are dominated
by 
zero-point motion,
we introduce a new 
method of scaling
which uses
\begin{eqnarray}
\label{eq:blur}
&&V^{\rm avg,\sigma}_{\rm Ne \mhyphen Ne}(|\vec{r}_1\!-\!\vec{r}_2|)
\!=\!\!
\int_{-\infty}^{\infty}\!\!\!\!\!\!\!\!dx_1' 
\int_{-\infty}^{\infty}\!\!\!\!\!\!\!\!dy_1' 
\int_{-\infty}^{\infty}\!\!\!\!\!\!\!\!dz_1' 
\int_{-\infty}^{\infty}\!\!\!\!\!\!\!\!dx_2' 
\int_{-\infty}^{\infty}\!\!\!\!\!\!\!\!dy_2' 
\int_{-\infty}^{\infty}\!\!\!\!\!\!\!\!dz_2' 
\nonumber
\\
&&\ \ \ \ \ \ \ \ \ \ 
n_\sigma(x_1')
n_\sigma(y_1')
n_\sigma(z_1')
n_\sigma(x_2')
n_\sigma(y_2')
n_\sigma(z_2')
\nonumber
\\
&&\ \ \ \ \ \ \ \ \ \ 
V_{\rm Ne \mhyphen Ne}
(|(\vec{r}_1+\pvec{r}_1')-(\vec{r}_2+\pvec{r}_2')|),
\end{eqnarray}
where 
$\pvec{r}_i'=(x'_i,y'_i,z'_i)$
and
$n_\sigma$
is a normal distribution with 
a mean of 
zero 
and a 
standard deviation 
of 
$\sigma$.
Eq.~(\ref{eq:blur})
calculates the 
interaction
of the two 
Ne
atoms,
each of which has a
three-dimensional 
gaussian 
distribution 
for its positional 
probability.
As in 
Eq.~(\ref{eq:scaled}),
it is necessary to scale 
this potential so that it
gives the correct value of 
$E_{\rm coh}$ 
and 
$b$.
This scaling can be achieved
by using either
\begin{equation}
V^{\rm mod2}_{\rm Ne \mhyphen Ne}(r)
=
V^{\rm avg,\sigma}(\beta' r)
\label{eq:mod2}
\end{equation}
or
\begin{equation}
V^{\rm mod3}_{\rm Ne \mhyphen Ne}(r)
=
\alpha' V^{\rm avg,\sigma'}(r).
\label{eq:mod3}
\end{equation}
For 
Eq.~(\ref{eq:mod2}), 
the constants required to match
the experimental values of 
$E_{\rm coh}$ 
and
$b$
are
$\sigma=0.197$~\AA~and
$\beta'=1.0348$.
For 
Eq.~(\ref{eq:mod3}), 
the parameters are
$\sigma'=0.145$~\AA~and
$\alpha'=0.853$.
The fact that 
$\alpha'$
and
$\beta'$
are both closer to unity 
than 
$\alpha$
and 
$\beta$
indicates that this 
averaging is 
more appropriate 
that the 
simple scaling of 
Eq.~(\ref{eq:scaled}).

The
BaF-Ne
interaction
energy,
$V_{\rm BaF \mhyphen Ne}(r,\theta)$,
of 
Eq.~(\ref{eq:totalE})
is obtained by interpolating 
the 
triatomic
results of 
Section~\ref{sec:triatomic}
that are listed in 
Table~\ref{table:sup0}
through
\ref{table:sup180}
of the 
Supplementary Materials.
When using the 
Ne-Ne 
potentials of 
Eqs.~(\ref{eq:mod2})
and 
(\ref{eq:mod3}),
the triatomic potential
$V_{\rm BaF \mhyphen Ne}(r,\theta)$
is also averaged over the 
three-dimensional
gaussian 
distributions for the 
Ne
position.
% by calculating 
% the 
% ground-state
% energies of the 
% BaF-Ar 
% triatomic
% system 
% for 
% 1386 
% values of 
% $r$
% and 
% $\theta$
% using 
% high-precision
% all-electron
% relativistic
% quantum-mechanical
% calculations 
% that
% include 
% correlation
% and that are 
% extrapolated to 
% the complete basis
% set limit.
% A fit provided in that work
% or,
% alternatively,
% interpolations
% and extrapolations
% provide 
% $V_{\rm BaF \mhyphen Ar}(r,\theta)$
% for intermediate values of 
% $r$
% and
% $\theta$.
% Uncertainties from 
% this calculation of 
% $V_{\rm BaF \mhyphen Ar}(r,\theta)$
% are also provided in 
% Ref.~\cite{koyanagi2023accurate}.

% ,
% and these uncertainties are used 
% in estimating the accuracy of these 
% of the present work.

% Four uncertainties in these
% simulations are investigated.
% The first is due to the 
% finite size of the cluster 
% used for the calculation. 
For all simulations, 
the calculations are repeated
with increasing numbers 
$N$ 
and 
$M$
of 
Ne
atoms
and
extrapolated to 
the limit of 
large
$N$
and
$M$.
The convergence of
our results with 
increasing cluster
size provides 
an estimate of the
resulting uncertainty.
Additionally, 
% A second comes from the 
% uncertainty in our 
% calculated
% BaF-Ne
% potentials. 
% in 
% Ref.~\cite{koyanagi2023accurate}.
% In that work, 
% we repeat our calculations
% with increasing basis set 
% sizes: 
% $n \zeta$,
% with 
% $n=2$
% through
% $5$.
% We make two extrapolations
% of our results,
% one from 
% $n=2$,
% $3$
% and
% $4$,
% and the other
% (more precise one)
% from 
% $n=3$,
% $4$
% and
% $5$.
% Based on comparisons to 
% measured quantities in 
% BaF, 
% Ba,
% Ba$^+$
% and
% Ar,
% we estimate the 
% uncertainty in our 
% 345
% extrapolated
% results
% to be
% one quarter
% of the difference 
% between these two 
% extrapolations.
% To determine the effect of 
% these uncertainties on our 
% calculations,
we repeat the simulations
with both the
234
and 
345
extrapolated
potentials from 
% Ref.~\cite{koyanagi2023accurate}.
Section~\ref{sec:triatomic}.
As discussed above, 
the
uncertainty in our simulations
is expected to be 
one quarter of the difference
between the 345 and 234 results. 
Finally,
% Thirdly, 
% and most importantly,
% we investigate the approximation
% inherent in
% using the scaled
% Ne-Ne
% potentials of 
% Eqs.~(\ref{eq:scaled}),
% (\ref{eq:mod2})
% and 
% (\ref{eq:mod3}).
% To do this, 
we repeat our simulations
using 
using the three 
modified 
Ne-Ne 
potentials
of
Eqs.~(\ref{eq:scaled}),
(\ref{eq:mod2})
and 
(\ref{eq:mod3}).
The variation 
of our results
using these three
potentials
gives a 
scale for the 
associated uncertainty.
Uncertainties due to 
four-atom
BaF-Ne-Ne
interactions
are expected to be 
very small, 
as was shown  
to be the case for 
similar 
BaF-Ar-Ar 
interactions
\cite{lambo2023calculation}.
% They should be 
% even smaller here
% given the 
% smaller binding 
% energies for the 
% Ne
% atom.
% the 
% $n=2$,
% $3$,
% $4$
% energies
% for the
% BaF-Ar-Ar
% four-atom system
% for twenty
% geometries
% with separations 
% of between
% 3
% and 
% 7~\AA,
% which covers the  
% most important range of 
% separations for
% our simulated solids.
% We 
% compare the 
% 234
% extrapolation of this 
% binding 
% energy 
% to the sum of the 
% Ar-Ar
% binding energy
% plus
% the two 
% BaF-Ar
% contributions
% (also calculated using a 
% 234 
% extrapolation).
% The difference 
% between the full
% BaF-Ar-Ar
% calculation
% and the sum of the 
% two-body 
% contributions
% is typically less than 
% one percent. 
% It 
% is sometimes positive 
% and sometimes negative,
% and gets smaller quickly
% with increasing 
% distances.
% Therefore, 
% we estimate the net effect
% due to 
% this 
% four-atom
% effect
% to be less than
% one percent.

% Calculations are repeated 
% with these coefficients set 
% to unity 
% to investigate the 
% sensitivity of our results to
% the form of 
% the 
% Ar-Ar
% potential.

\section{
Results
}
\subsection{
Number of  
%A\MakeLowercase{r}
Ne
atoms substituted
for 
a 
%B\MakeLowercase{a}F
BaF
molecule
}
% When embedded in a matrix,
% the 
% BaF
% molecule 
% substitutes for 
% $S$
% Ne
% atoms.
To determine 
which integer
$S$
of 
Ne
atoms 
displaced by
the 
BaF
molecule
is most energetically
favourable,
we compare
values of 
\begin{equation}
\Delta E_{n,m,S}
=
E_{n,m,S}-S E_{\rm coh},
\label{eq:EvsS}
\end{equation}
where 
$E_{n,m,S}$
is calculated  
using 
Eq.~(\ref{eq:totalE})
for 
a sphere of 
radius 
$b_m$
of 
non-fixed
Ne
atoms
(of which 
$S$
are removed and 
replaced with 
a
BaF
molecule)
inside of a  
spherical shell
of fixed 
Ne
atoms
that
extends to 
a radius of 
$b_n$.
The 
$S E_{\rm coh}$
term
corrects for
the missing 
Ne
cohesive
energy 
from removing
$S$
isolated
Ne 
atoms.
The lowest 
value of 
$\Delta E_{n,m,S}$
occurs for 
$S=10$.
This is in contrast 
to the case for 
a 
BaF
molecule in 
an 
Ar
solid
\cite{lambo2023calculation}, 
where 
the much tighter
binding leads to
$S=4$
being preferred,
but is similar 
to the case 
of an 
Yb 
atom in a
Ne
solid
\cite{lambo2021high}, 
which also
prefers
$S=10$.

% a
% neutral 
% Ba
% atom,
% where 
% $S=6$
% is the preferred 
% substitution 
% \cite{kleshchina2019stable}.
% As can be seen in 
% Ref.~\cite{koyanagi2023accurate},
% the 
% F
% side of the 
% BaF
% molecule
% bonds more strongly
% to the
% Ar 
% atoms
% and this reduces 
% the preferred
% value of
% $S$
% for 
% BaF 
% as 
% compared 
% to 
% Ba.

\begin{figure}
\includegraphics
[width=0.9\linewidth]{
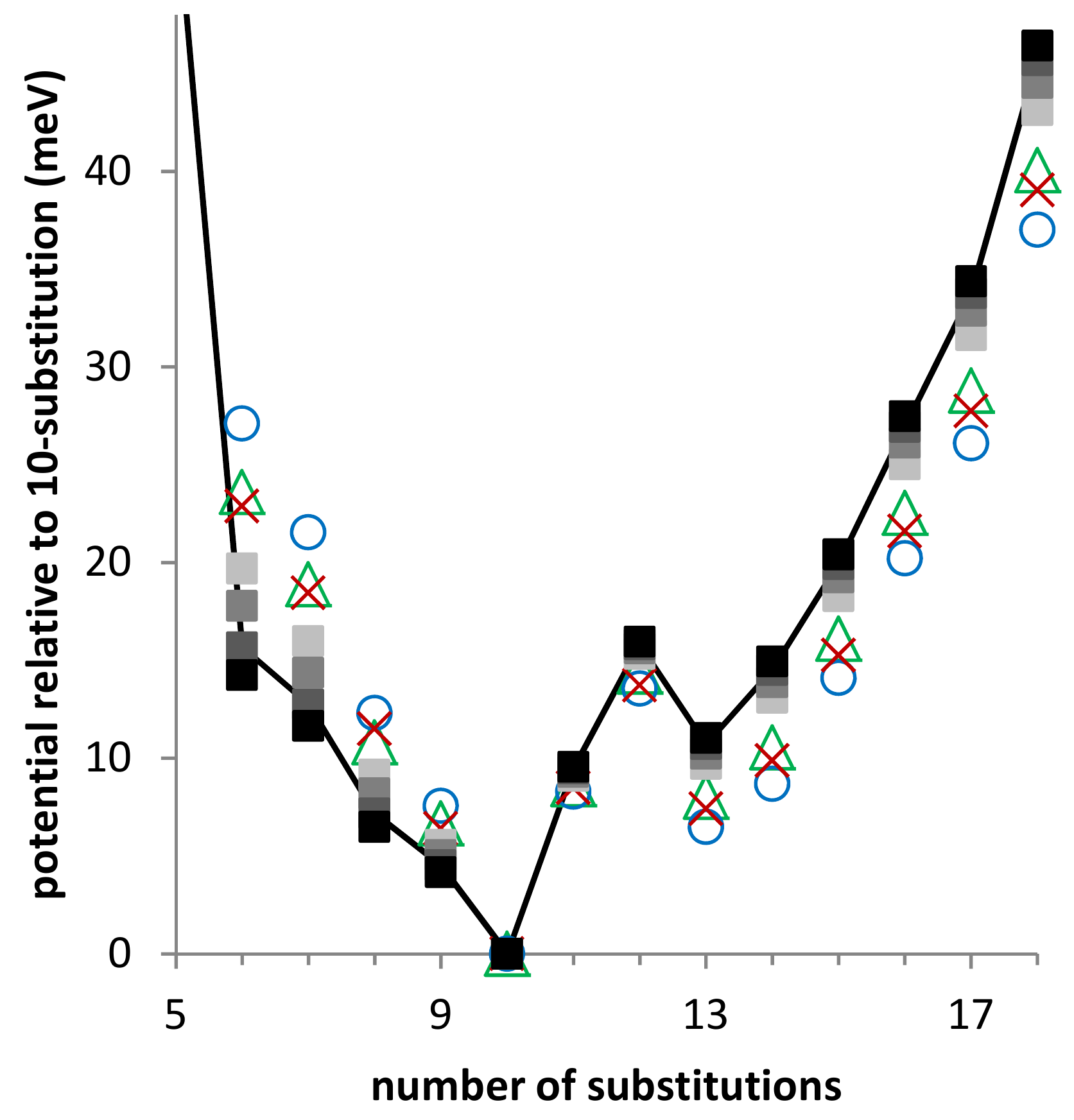}
\caption{
(color online)
Energy favourability
versus
the number
$S$
of
Ne
atoms substituted
for a 
BaF 
molecule.
The data plotted is 
that of 
Table~\ref{table:subs}.
Squares of increasing 
darkness represent the 
simulations 
with increasing numbers
of
Ne 
atoms, 
with the black squares
showing the 
extrapolation to
a macroscopic
matrix.
The 
exes
and 
open 
symbols
correspond to
the rows marked
$a$, 
$b$
and 
$c$
in 
Table~\ref{table:subs},
and
(when 
compared 
to the 
lightest
gray squares)
show 
the sensitivity 
to
the choice of 
BaF-Ne
potential
and 
Ne-Ne
potential.
$S=10$
is favoured, 
with a second minimum
present at 
$S=13$.
}
\label{fig:subs}
\end{figure}

% \begin{figure}
% \includegraphics
% [width=1.0\linewidth]{
% BaFNeContourPlot}
% \caption{
% (color online)
% .
% }
% \label{fig:contours}
%\end{figure}

\begin{table*}%[hbt!]
\begin{ruledtabular}
\caption{\label{table:subs} 
Comparison of energetic 
favourability
of a
BaF
molecule substituting for 
$S$
Ne 
atoms
for increasing 
cluster size
$N$
with an
increasing
number 
$M$
of
non-fixed
Ne 
atoms.
To aid in determining
the uncertainties of the 
simulations, 
calculations using 
a less precise 
234 
form of the 
BaF-Ne 
potential
are also shown,
as are 
calculations 
for which the 
Ne-Ne
potential is 
obtained from 
Eqs.~(\ref{eq:mod2})
and 
(\ref{eq:mod3}).
$S=10$
is favoured, 
with a second local 
minimum at 
$S=13$.
}
\begin{tabular}{ccccccccccccccccccc}
&&&&
\multicolumn{14}{c}{$
\Delta E_{n,m,S}
-
\Delta E_{n,m,S=10}
$ (meV)}\\
$n$
&
$m$
&
$N_{S=10}$
&
$M_{S=10}$
&
$S$: 5
&
6
&
7
&
8
&
9
&
10
&
11
&
12
&
13
&
14
&
15
&
16
&
17
&
18
\\
\hline
% 13
% &
% 5
% &
% 311
% &
% 69
% &
% 73
% &
% 24
% &
% 20
% &
% 12
% &
% 7
% &
% 0
% &
% 9
% &
% 15
% &
% 9
% &
% 13
% &
% 17
% &
% 23
% &
% 30
% &
% 41
% \\
19
&
7
&
521
&
125
&
64
&
20
&
16
&
9
&
6
&
0
&
9
&
15
&
10
&
13
&
18
&
25
&
32
&
43
\\
19
&
7
&
521
&
125
&
70$^a$
&
23$^a$
&
18$^a$
&
11$^a$
&
6$^a$
&
0$^a$
&
8$^a$
&
14$^a$
&
7$^a$
&
10$^a$
&
15$^a$
&
22$^a$
&
28$^a$
&
39$^a$
\\
19
&
7
&
521
&
125
&
71$^b$
&
24$^b$
&
19$^b$
&
11$^b$
&
7$^b$
&
0$^b$
&
9$^b$
&
14$^b$
&
8$^b$
&
11$^b$
&
16$^b$
&
23$^b$
&
29$^b$
&
40$^b$
\\
19
&
7
&
521
&
125
&
77$^c$
&
27$^c$
&
22$^c$
&
12$^c$
&
8$^c$
&
0$^c$
&
8$^c$
&
14$^c$
&
6$^c$
&
9$^c$
&
14$^c$
&
20$^c$
&
26$^c$
&
37$^c$
\\
27
&
11
&
877
&
215 
&
59
&
18
&
14
&
8
&
5
&
0
&
9
&
15
&
10
&
13
&
19
&
26
&
33
&
45
\\
47
&
20
&
1951 
&
545
&
54
&
16
&
13
&
7
&
5
&
0
&
9
&
16
&
11
&
15
&
20
&
27
&
34
&
46
\\ 
\multicolumn{4}{c}{extrapolated}
&
51
&
14
&
12
&
6
&
4
&
0
&
10
&
16
&
11
&
15
&
20
&
27
&
34
&
46
\\
\end{tabular}
\end{ruledtabular}
%\begin{tablenotes}
$^a$This row uses the 
less-precise 
234 
BaF-Ne 
potential. 
One quarter of the difference between this
entry and the 
entry above it 
gives an estimate of 
the uncertainty
for the 
previous row.
% due to uncertainties in the potential calculated
% in 
% Ref.~\cite{koyanagi2023accurate}.
\\
$^b$This row uses the Ne-Ne potential
of 
Eq.~(\ref{eq:mod2}). 
\\
$^c$This row uses the Ne-Ne potential
of 
Eq.~(\ref{eq:mod3}). 
The variation between this row,
the previous row and the first row
provides a scale for the approximation 
implicit in 
Eq.~(\ref{eq:scaled}).
%\end{tablenotes}
\end{table*}

Table~\ref{table:subs}
and 
Fig.~\ref{fig:subs}
show the values
of 
$
\Delta E_{n,m,S}
-
\Delta E_{n,m,S=10}
$
for various values 
of 
$n$
and
$m$.
As can be seen 
from the table,
substituting a 
BaF
molecule 
for 
$S=10$
Ne
atoms 
is energetically 
favourable compared to 
other values 
of 
$S$.
The 
$\sim4$~meV
energy advantage of the 
$S=10$ substitution
is much larger than the 
thermal energy scale 
$k_B T=0.34$~meV
for a matrix
held at 
4~kelvin. 
As a result, 
it can be expected
that the
BaF 
molecules will
persist in an
$S=10$
site 
at 
this temperature.
As seen from 
Fig.~\ref{fig:subs},
a second (local) minimum
occurs for   
$S=13$,
and this second minimum
also is favoured by 
$\sim$4~meV
compared to neighbouring 
values of 
$S$.

The conclusions drawn from 
Table~\ref{table:subs}
are not affected by 
any of the uncertainties
that we investigated,
as illustrated in 
Fig.\ref{fig:subs}.
It was possible to 
extrapolate
$E_{n,m,S}$
to 
its large 
$N$
and
$M$
limit,
$E_{{\rm extrap},S}$,
by noting that  
$E_{{\rm extrap},S}-E_{n,m,S}$
scales as 
$1/M$
(so long as 
$N$ 
is approximately 
proportional to
$M$).
The extrapolated values
are given in the final 
row of 
Table~\ref{table:subs}, 
and the uncertainty in these 
extrapolations is less 
than 
1~meV.
From 
Fig.~\ref{fig:subs},
it can be seen that 
using the less precise
234 
extrapolation for 
the 
BaF-Ne
interaction energy
(shown as exes in the figure)
does not affect these 
conclusions.
Similarly, 
the use of alternate 
scaling methods for the 
Ne-Ne potential
(Eqs.~(\ref{eq:mod2})
and 
(\ref{eq:mod3}) --
shown as triangles
and circles
in the figure)
also do not affect
the conclusions. 

\subsection{
Geometry of the matrix
isolated BaF
\label{subSect:geom}}

\begin{figure}
\includegraphics
[width=0.9\linewidth]{
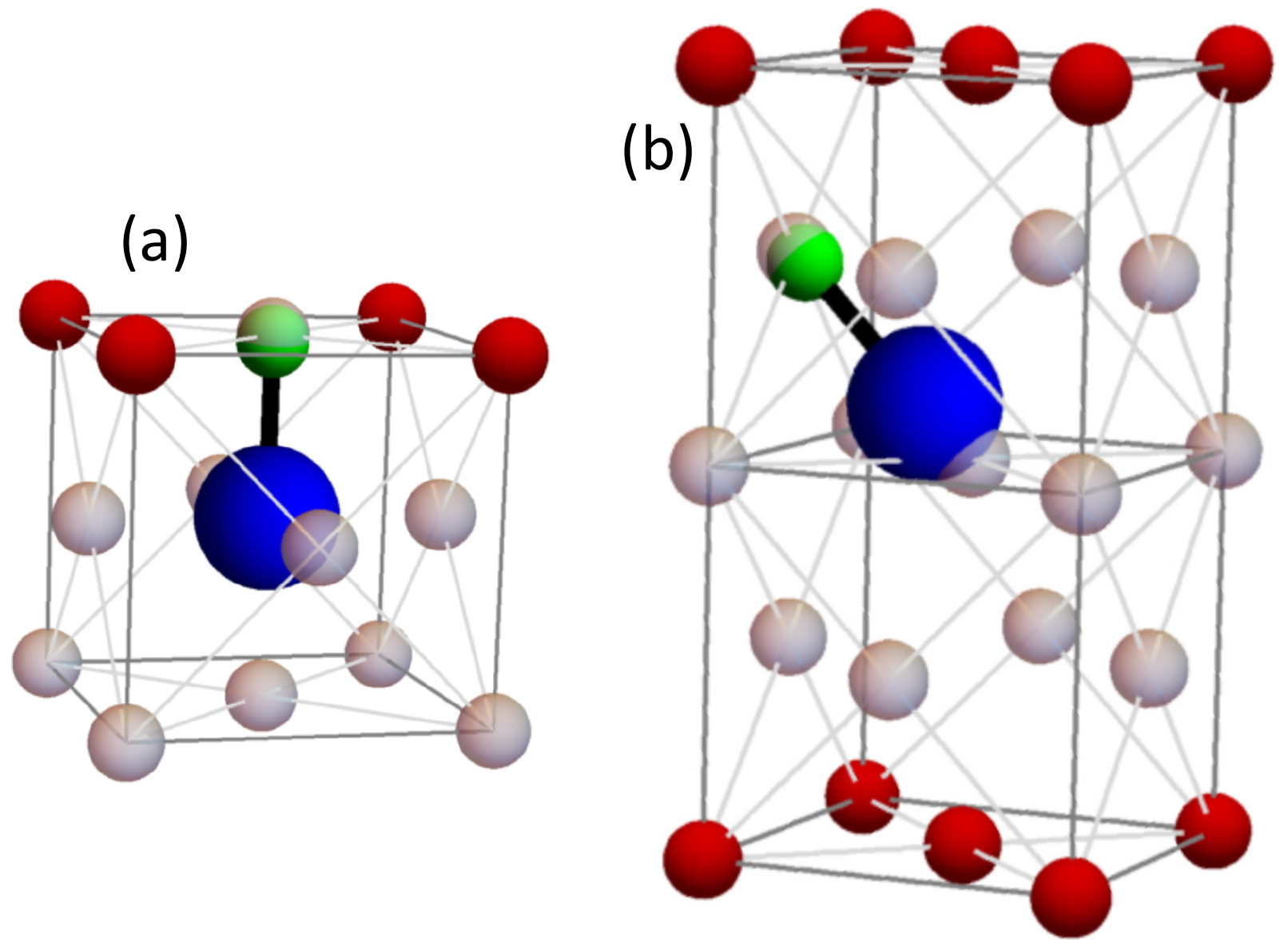}
\caption{
(color online)
The 
most energetically 
favourable configuration
(a)
has the 
BaF
molecule 
substituting for 
ten
Ne 
atoms
(shown in grey)
and aligned along the 
$\langle 100 \rangle$
axis 
of the 
Ne
fcc
crystal.
The 
F
atom 
(green)
is situated 
near one of the 
substituted 
Ne 
atoms
and the 
Ba
atom 
(blue)
is situated near the 
centre of the 
fcc 
cube. 
Only four of the 
fourteen
Ne
atoms 
(red)
remain on 
the surface of 
this cube.
Panel 
(b)
shows
the geometry for
the 
BaF
molecule substituting for 
thirteen 
Ne atoms.
In this case, 
a 
Ne
atom and its
twelve
nearest 
neighbours are 
replaced by the 
BaF 
molecule,
with the 
Ba
atom situated
near the central
missing 
Ne
atom
and the 
F
atom
situated near 
one of the nearest 
neighbours.
The 
BaF
axis is along 
the 
110
direction of the 
Ne 
crystal for this case.
}
\label{fig:s10Geom}
\end{figure}

At its minimum 
energy the favoured
$S=10$ 
substitution
has the 
BaF
molecule
aligned with the 
$\langle 100 \rangle$
axis of the 
Ne
fcc
crystal, 
as shown in 
Fig.~\ref{fig:s10Geom}(a),
where the 
F 
atom replaces the 
Ne
atom
at the top face 
of a cube,
with the 
Ba
atom located near the 
centre of this
cube.
The 
four 
Ne
atoms at the four
top corners of 
the cube 
tightly bind to
the 
F
atom,
whereas the other
Ne
atoms on this cube
are displaced by the 
larger
Ba
atom. 
The 
original 
fcc
crystal positions of the 
ten
missing 
Ne
atoms 
are shown in 
grey in the figure. 

The positions of the 
BaF
molecule
and the nearest
Ne
atom 
neighbours
(for 
$S=10$)
are 
detailed in 
Table~\ref{table:BaFNeposition}.
The equilibrium positions of the 
Ne 
atoms are remarkably close
to their original positions
in the 
fcc crystal.
The uncertainties for the 
Ne
atom
positions
in the table
are
approximately 
0.01~\AA~or less,
as was demonstrated
by repeating the 
simulations with larger
clusters,
with the 
less-precise
$E_{234}$
potential,
and 
with the modified
potentials of 
Eqs.~\ref{eq:mod2}
and
\ref{eq:mod3}.
The displacement
of the 
BaF 
molecule
along the 
$\langle 100 \rangle$
axis
($x$ 
in
Table~\ref{table:BaFNeposition})
shows some dependence on the 
form of the 
Ne-Ne 
potential used
and therefore
has an uncertainty 
of 
0.05~\AA. 

The configuration for the 
local minimum at
$S=13$
is shown in 
Fig.~\ref{fig:s10Geom}(b).
For this case, 
the 
BaF
molecule is parallel to the 
110
axis of the crystal.
The 
Ba
atom 
replaces the central
Ne
atom
and its 
twelve
nearest
neighbours,
with the
F
atom 
nearly at the 
location of one 
of these 
nearest neighbours.
The detailed positions of the 
BaF 
molecule
and the neighbouring 
Ne 
atoms for 
$S=13$
are given in 
Table~\ref{table:S13positions}
of the 
Supplementary Materials.

\begin{table}%[hbt]
\begin{ruledtabular}
\caption{\label{table:BaFNeposition} 
The
equilibrium positions
(in \AA~relative
to the
centre 
of the cluster)
for the 
lowest-energy
($S=10$)
configuration
of a 
BaF
molecule 
and 
its
nearby
Ne
atom
neighbours.
Each row represents
$N_{\rm symm}$
symmetrical atoms
(using the symmetries
about 
$z=0$,
$y=0$
and
$y=z$).
The final three columns
give the displacement of the
F 
and 
Ne
atoms 
relative to the ideal
fcc positions.}
\begin{tabular}{cccccccc}
&
$N_{\rm symm}$&
$x$&
$y$&
$z$&
$\Delta_x$&
$\Delta_y$&
$\Delta_z$\\
\hline
Ba& &-2.24&0.00&0.00                         \\
F & &-0.08&0.00&0.00&-0.08& 0.00& 0.00\\
Ne&4& 0.11&2.31&2.31& 0.11& 0.08& 0.08\\
Ne&4& 2.23&2.22&0.00& 0.00&-0.01& 0.00\\
Ne&4& 0.01&4.45&0.00& 0.01&-0.01& 0.00\\
Ne&1& 4.45&0.00&0.00&-0.01& 0.00& 0.00\\
Ne&8&-2.22&2.25&4.50& 0.01& 0.02& 0.03\\
Ne&8& 2.27&2.24&4.49& 0.03& 0.01& 0.03\\
Ne&4& 4.46&2.23&2.23&-0.01& 0.00& 0.00\\
Ne&4&-4.49&4.53&0.00&-0.03& 0.07& 0.00\\
Ne&4& 0.02&4.49&4.49& 0.02& 0.03& 0.03\\
Ne&4& 4.47&4.46&0.00& 0.00& 0.00& 0.00\\
Ne&4&-6.79&2.26&0.00&-0.10& 0.03& 0.00\\
Ne&4&-2.22&6.71&0.00& 0.01& 0.02& 0.00\\
Ne&8& 0.01&6.70&0.00& 0.01& 0.00& 0.00\\
Ne&4& 2.24&6.70&0.00& 0.01& 0.01& 0.00\\
Ne&4& 6.69&2.23&0.00&-0.01& 0.00& 0.00\\
\end{tabular}
\end{ruledtabular}
\end{table}

\subsection{
Potential for preventing
BaF rotations and migration}

\begin{figure}
\includegraphics[width=1.0\linewidth]{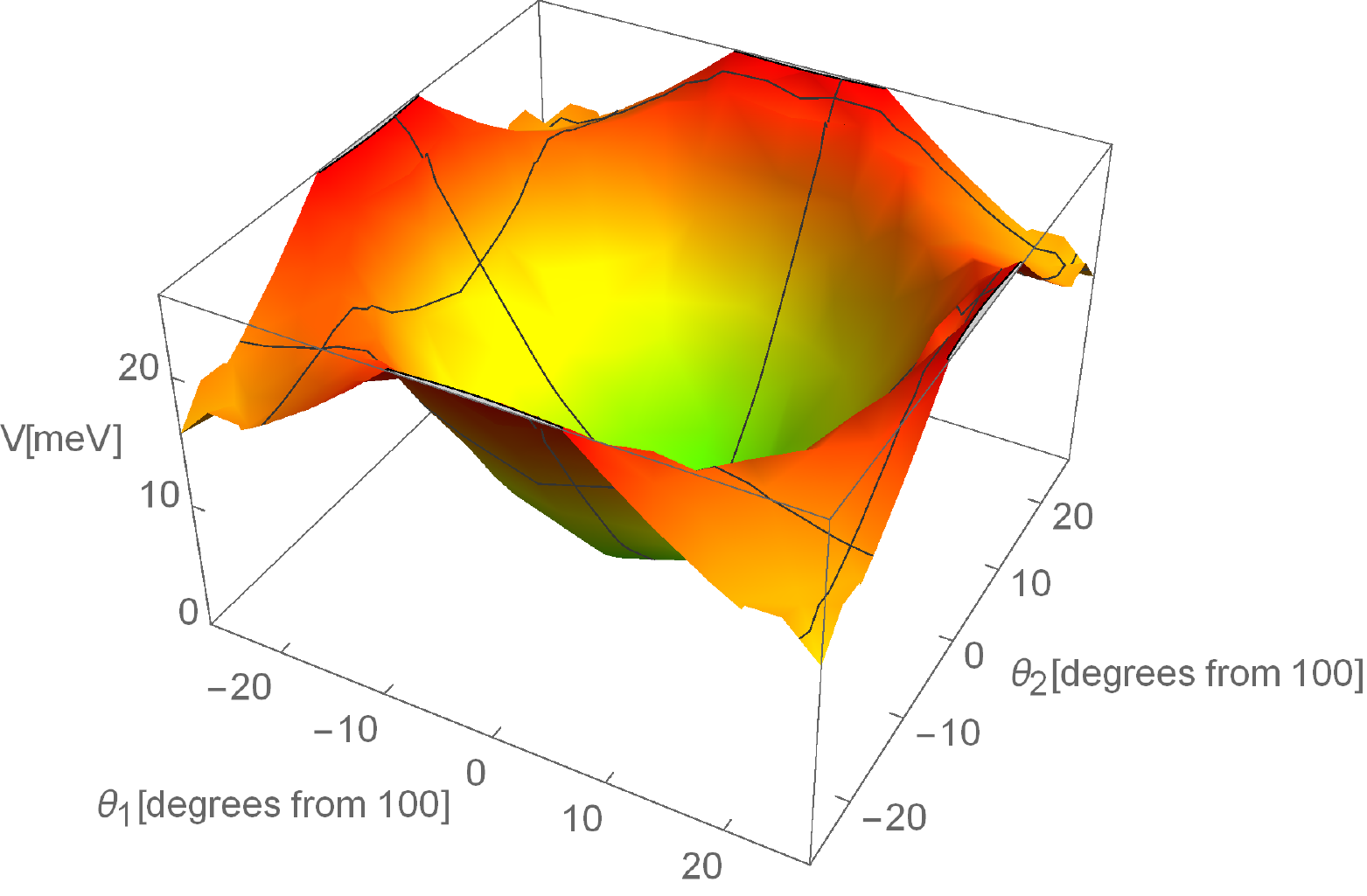}
\caption{
(color online)
$V(\theta_1,\theta_2)$ 
shows the
energy cost 
in meV
for changing the 
orientation of the 
BaF
molecule from
its preferred
$\langle 100 \rangle$
axis
in two perpendicular 
directions:
$\theta_1$
(towards the 
$\langle 010 \rangle$
direction)
and 
$\theta_2$
(towards the 
$\langle 001 \rangle$
direction).
}
\label{fig:angleBarrier}
\end{figure}

The 
potential energy barrier
confining the 
angular orientation of the 
BaF
molecule 
to near the 
$\langle 100 \rangle$
axis of the 
Ne
crystal
is calculated
by repeating 
the energy optimization
for many fixed orientations 
of the 
BaF
molecule. 
To do this, 
the energy minimization 
for the cluster
is performed
while varying the remaining 
$3M+3$
parameters
(the locations of the 
molecule and the 
$M$
Ne 
atoms), 
with the orientation of the 
molecule fixed.
The resulting 
potential barrier
is shown in 
Fig.~\ref{fig:angleBarrier}.
From the figure, 
it can be seen that 
a 
20-meV-deep
potential
keeps the 
BaF
molecule aligned along 
the 
$\langle 100 \rangle$
axis.
Identical wells are present
at six symmetric 
axes: 
$\langle \pm100 \rangle$,
$\langle 0\pm10 \rangle$
and
$\langle 00\pm1 \rangle$.
At 
4~kelvin,
the
20-meV
wells are 
 more than sufficient to 
confine the orientation
to one of these 
six
orientations.
The locked orientation
(together with methods to 
separately address
individual orientations
\cite{vutha2018orientation})
makes it unnecessary 
to apply an electric field
during an 
eEDM 
measurement.

Similarly,
to determine the 
energy cost for a 
BaF 
molecule being
at a position
away 
from its 
equilibrium,
the simulations are repeated
with the 
BaF
centre-of-mass
position 
fixed
at many displacements 
from equilibrium,
and 
minimizing the energy by 
varying the remaining
$3M+2$
parameters.
These simulations 
indicate that the 
energy cost for moving 
by a distance of 
0.7~\AA~from
equilibrium
in any direction
is
20~meV
or greater.
As a result, 
the 
BaF
molecule is 
unable to 
migrate 
through the solid.

\section{
conclusions}
Calculations 
of the local environment 
of a 
BaF
molecule within
a Ne matrix
are discussed.
It is found that
the molecule
prefers to
replace 
ten
Ne
atoms
and 
is aligned along the 
$\langle 100 \rangle$ 
axis of 
the 
Ne crystal.
The remaining 
Ne
atoms are 
found to be only 
slightly 
displaced 
from their 
original 
fcc
crystal positions.
The inability of
the molecule
to reorient or 
migrate at 
a temperature of
4 kelvin 
and 
the small perturbation
of the rest of the 
Ne
crystal 
are important 
features
for the 
planned 
eEDM
measurement 
by the 
EDM$^3$
collaboration 
using 
matrix-isolated
BaF
molecules.

\section{acknowledgements}
This work is supported by
the
Alfred P. Sloan Foundation,
the 
Gordon and Betty Moore Foundation,
the 
Templeton Foundation
in conjunction with the 
Northwestern Center for Fundamental Physics,
the
Natural Sciences and Engineering Research Council
of Canada
and
York University.
Computations for this work were 
supported by
Compute Ontario 
and 
the Digital Research Alliance of Canada.

\newpage
\section{Supplementary Materials}
Tables~\ref{table:sup0}
to
\ref{table:sup180}
show the 
calculated 
potential energies 
$V(r,\theta)$ 
(in meV) 
for the basis sets 
$\zeta=$
2 through 5
for angles 
$\theta$
given in the 
table captions.
Also shown 
are the 
complete-basis-set
extrapolations of 
Eqs.~(\ref{eq:234}) 
and 
(\ref{eq:345}).

Values in italics
are 
interpolations 
or extrapolations.
The italics values 
for 
$r<10$~\AA~are 
based on the 
$E_{234}$ 
values
plus 
an interpolation 
of 
$E_{345}-E_{234}$.
For
$r>10$~\AA,
the values in italics
are 
a smooth
extrapolation 
of  
intermediate-range
data
using a fit
to inverse powers
of 
$r$.

\begin{table}
\begin{ruledtabular}
\setlength\extrarowheight{-2pt}
\scriptsize
\caption{
Potential energy $E(r,\theta)$ in meV for the basis sets 
$\zeta=$ 2 through 5 and for the
complete-basis-set extrapolations of
Eqs.~(\ref{eq:234}) and (\ref{eq:345})
 for $\theta=$
0
$^\circ$.
}
\label{table:sup0}
\begin{tabular}
{rrrrrrrr}
$r$&
$E(2\zeta)$&
$E(3\zeta)$&
$E(4\zeta)$&
$E(5\zeta)$&
$E(234)$&
$E(345)$\\
\hline
2.4&10\,350.1&10\,004.2&9\,864.4&9\,812.2&9\,781.9&9\,781.8\\
2.5&6\,828.2&6\,565.6&6\,461.5&6\,422.3&6\,400.2&6\,399.5\\
2.6&4\,476.5&4\,285.6&4\,206.2&---&4\,159.1&\textit{4\,160.1}\\
2.7&2\,925.5&2\,789.9&2\,728.5&2\,706.9&2\,691.8&2\,694.4\\
2.8&1\,908.4&1\,811.9&1\,764.5&---&1\,735.8&\textit{1\,738.5}\\
2.9&1\,243.0&1\,173.1&1\,136.8&1\,124.5&1\,114.7&1\,117.4\\
3.0&808.1&756.0&728.5&---&711.8&\textit{713.6}\\
\end{tabular}
\end{ruledtabular}
\end{table}
\begin{table}
\begin{ruledtabular}
\setlength\extrarowheight{-2pt}
\scriptsize
\caption{
Potential energy $E(r,\theta)$ in meV for the basis sets 
$\zeta=$ 2 through 5 and for the
complete-basis-set extrapolations of
Eqs.~(\ref{eq:234}) and (\ref{eq:345})
 for $\theta=$
10
$^\circ$.
}
\label{table:sup10}
\begin{tabular}
{rrrrrrrr}
$r$&
$E(2\zeta)$&
$E(3\zeta)$&
$E(4\zeta)$&
$E(5\zeta)$&
$E(234)$&
$E(345)$\\
\hline
2.4&9\,166.9&8\,845.5&8\,717.8&8\,669.6&8\,642.5&8\,641.6\\
2.5&6\,059.9&5\,819.2&5\,723.0&5\,687.0&5\,666.3&5\,666.1\\
2.6&3\,983.5&3\,809.8&3\,735.7&---&3\,691.7&\textit{3\,693.1}\\
2.7&2\,610.5&2\,487.0&2\,429.6&2\,409.6&2\,395.0&2\,398.0\\
2.8&1\,707.4&1\,619.0&1\,574.6&---&1\,547.6&\textit{1\,550.4}\\
2.9&1\,114.7&1\,050.0&1\,016.0&1\,004.6&995.3&997.9\\
3.0&726.0&677.5&651.8&---&636.1&\textit{637.8}\\
\end{tabular}
\end{ruledtabular}
\end{table}
\begin{table}
\begin{ruledtabular}
\setlength\extrarowheight{-2pt}
\scriptsize
\caption{
Potential energy $E(r,\theta)$ in meV for the basis sets 
$\zeta=$ 2 through 5 and for the
complete-basis-set extrapolations of
Eqs.~(\ref{eq:234}) and (\ref{eq:345})
 for $\theta=$
20
$^\circ$.
}
\label{table:sup20}
\begin{tabular}
{rrrrrrrr}
$r$&
$E(2\zeta)$&
$E(3\zeta)$&
$E(4\zeta)$&
$E(5\zeta)$&
$E(234)$&
$E(345)$\\
\hline
2.4&6\,482.1&6\,230.3&6\,130.0&6\,092.4&6\,071.0&6\,070.5\\
2.5&4\,311.8&4\,128.0&4\,050.2&4\,022.0&4\,004.0&4\,005.6\\
2.6&2\,855.1&2\,723.4&2\,662.5&---&2\,625.9&\textit{2\,628.4}\\
2.7&1\,884.9&1\,790.2&1\,742.7&1\,726.6&1\,713.8&1\,717.2\\
2.8&1\,241.3&1\,171.9&1\,135.2&---&1\,112.7&\textit{1\,115.5}\\
2.9&815.3&763.1&735.1&725.5&718.0&719.9\\
3.0&533.6&493.4&472.2&---&459.3&\textit{460.5}\\
\end{tabular}
\end{ruledtabular}
\end{table}
\begin{table}
\begin{ruledtabular}
\setlength\extrarowheight{-2pt}
\scriptsize
\caption{
Potential energy $E(r,\theta)$ in meV for the basis sets 
$\zeta=$ 2 through 5 and for the
complete-basis-set extrapolations of
Eqs.~(\ref{eq:234}) and (\ref{eq:345})
 for $\theta=$
30
$^\circ$.
}
\label{table:sup30}
\begin{tabular}
{rrrrrrrr}
$r$&
$E(2\zeta)$&
$E(3\zeta)$&
$E(4\zeta)$&
$E(5\zeta)$&
$E(234)$&
$E(345)$\\
\hline
2.4&3\,834.6&3\,670.2&3\,598.1&3\,572.2&3\,555.0&3\,557.2\\
2.5&2\,575.1&2\,455.9&2\,398.7&2\,378.9&2\,364.1&2\,367.4\\
2.6&1\,722.6&1\,635.4&1\,590.3&---&1\,562.7&\textit{1\,565.8}\\
2.7&1\,148.4&1\,083.4&1\,048.2&1\,036.4&1\,026.7&1\,029.5\\
2.8&763.0&713.3&686.2&---&669.7&\textit{671.5}\\
2.9&505.0&466.0&445.4&437.9&432.8&433.6\\
3.0&332.6&301.4&285.7&---&276.1&\textit{276.4}\\
\end{tabular}
\end{ruledtabular}
\end{table}
\begin{table}
\begin{ruledtabular}
\setlength\extrarowheight{-2pt}
\scriptsize
\caption{
Potential energy $E(r,\theta)$ in meV for the basis sets 
$\zeta=$ 2 through 5 and for the
complete-basis-set extrapolations of
Eqs.~(\ref{eq:234}) and (\ref{eq:345})
 for $\theta=$
40
$^\circ$.
}
\label{table:sup40}
\begin{tabular}
{rrrrrrrr}
$r$&
$E(2\zeta)$&
$E(3\zeta)$&
$E(4\zeta)$&
$E(5\zeta)$&
$E(234)$&
$E(345)$\\
\hline
2.4&2\,017.9&1\,921.2&1\,871.8&1\,855.0&1\,841.7&1\,845.2\\
2.5&1\,370.3&1\,297.4&1\,258.2&1\,244.9&1\,234.2&1\,237.3\\
2.6&926.7&870.6&839.8&---&821.0&\textit{823.1}\\
2.7&624.2&579.9&556.1&547.6&541.5&542.7\\
2.8&418.5&382.9&364.5&---&353.3&\textit{353.8}\\
2.9&279.2&250.1&235.8&230.3&227.2&227.1\\
3.0&185.2&161.0&149.8&---&143.1&\textit{142.9}\\
\end{tabular}
\end{ruledtabular}
\end{table}
\begin{table}
\begin{ruledtabular}
\setlength\extrarowheight{-2pt}
\scriptsize
\caption{
Potential energy $E(r,\theta)$ in meV for the basis sets 
$\zeta=$ 2 through 5 and for the
complete-basis-set extrapolations of
Eqs.~(\ref{eq:234}) and (\ref{eq:345})
 for $\theta=$
50
$^\circ$.
}
\label{table:sup50}
\begin{tabular}
{rrrrrrrr}
$r$&
$E(2\zeta)$&
$E(3\zeta)$&
$E(4\zeta)$&
$E(5\zeta)$&
$E(234)$&
$E(345)$\\
\hline
2.4&1\,007.5&946.9&914.1&902.6&893.9&896.0\\
2.5&691.9&643.5&617.5&608.3&601.7&602.9\\
2.6&473.2&433.7&413.3&---&400.9&\textit{401.6}\\
2.7&322.3&289.6&273.5&267.5&263.8&263.9\\
2.8&218.5&191.0&178.3&---&170.7&\textit{170.7}\\
2.9&147.3&124.1&113.9&110.0&107.8&107.7\\
3.0&98.58&78.94&70.70&---&65.81&\textit{65.71}\\
\end{tabular}
\end{ruledtabular}
\end{table}
\begin{table}
\begin{ruledtabular}
\setlength\extrarowheight{-2pt}
\scriptsize
\caption{
Potential energy $E(r,\theta)$ in meV for the basis sets 
$\zeta=$ 2 through 5 and for the
complete-basis-set extrapolations of
Eqs.~(\ref{eq:234}) and (\ref{eq:345})
 for $\theta=$
55
$^\circ$.
}
\label{table:sup55}
\begin{tabular}
{rrrrrrrr}
$r$&
$E(2\zeta)$&
$E(3\zeta)$&
$E(4\zeta)$&
$E(5\zeta)$&
$E(234)$&
$E(345)$\\
\hline
2.4&713.0&661.9&634.7&624.8&618.1&619.1\\
2.5&492.7&450.7&429.2&421.2&416.1&416.6\\
2.6&339.3&304.3&287.2&---&276.8&\textit{277.1}\\
2.7&232.8&203.3&189.6&184.4&181.4&181.4\\
2.8&159.0&134.0&123.0&---&116.4&\textit{116.5}\\
2.9&108.01&86.79&77.84&74.49&72.52&72.55\\
3.0&72.77&54.86&47.49&---&43.13&\textit{43.20}\\
\end{tabular}
\end{ruledtabular}
\end{table}
\begin{table}
\begin{ruledtabular}
\setlength\extrarowheight{-2pt}
\scriptsize
\caption{
Potential energy $E(r,\theta)$ in meV for the basis sets 
$\zeta=$ 2 through 5 and for the
complete-basis-set extrapolations of
Eqs.~(\ref{eq:234}) and (\ref{eq:345})
 for $\theta=$
60
$^\circ$.
}
\label{table:sup60}
\begin{tabular}
{rrrrrrrr}
$r$&
$E(2\zeta)$&
$E(3\zeta)$&
$E(4\zeta)$&
$E(5\zeta)$&
$E(234)$&
$E(345)$\\
\hline
2.4&512.7&467.3&444.0&435.1&429.8&430.0\\
2.5&356.9&318.9&300.3&293.2&289.0&289.1\\
2.6&247.9&215.8&200.8&---&191.8&\textit{191.8}\\
2.7&171.7&144.4&132.3&127.7&125.0&125.1\\
2.8&118.54&95.31&85.39&---&79.48&\textit{79.64}\\
2.9&81.41&61.75&53.56&50.63&48.70&48.93\\
3.0&55.47&38.97&32.15&---&28.11&\textit{28.36}\\
\end{tabular}
\end{ruledtabular}
\end{table}
\begin{table}
\begin{ruledtabular}
\setlength\extrarowheight{-2pt}
\scriptsize
\caption{
Potential energy $E(r,\theta)$ in meV for the basis sets 
$\zeta=$ 2 through 5 and for the
complete-basis-set extrapolations of
Eqs.~(\ref{eq:234}) and (\ref{eq:345})
 for $\theta=$
65
$^\circ$.
}
\label{table:sup65}
\begin{tabular}
{rrrrrrrr}
$r$&
$E(2\zeta)$&
$E(3\zeta)$&
$E(4\zeta)$&
$E(5\zeta)$&
$E(234)$&
$E(345)$\\
\hline
2.4&380.9&338.6&317.5&309.1&304.7&304.3\\
2.5&267.7&231.8&214.8&208.2&204.5&204.4\\
2.6&188.0&157.5&143.7&---&135.4&\textit{135.4}\\
2.7&131.96&106.04&94.68&90.53&87.89&88.12\\
2.8&92.48&70.50&61.11&---&55.52&\textit{55.84}\\
2.9&64.60&46.11&38.28&35.60&33.63&34.04\\
3.0&44.89&29.45&22.88&---&18.98&\textit{19.39}\\
\end{tabular}
\end{ruledtabular}
\end{table}
\begin{table}
\begin{ruledtabular}
\setlength\extrarowheight{-2pt}
\scriptsize
\caption{
Potential energy $E(r,\theta)$ in meV for the basis sets 
$\zeta=$ 2 through 5 and for the
complete-basis-set extrapolations of
Eqs.~(\ref{eq:234}) and (\ref{eq:345})
 for $\theta=$
70
$^\circ$.
}
\label{table:sup70}
\begin{tabular}
{rrrrrrrr}
$r$&
$E(2\zeta)$&
$E(3\zeta)$&
$E(4\zeta)$&
$E(5\zeta)$&
$E(234)$&
$E(345)$\\
\hline
2.4&299.3&258.2&237.7&229.4&225.3&224.6\\
2.5&212.8&177.9&161.2&154.7&151.2&151.0\\
2.6&151.6&122.0&108.4&---&100.2&\textit{100.2}\\
2.7&108.31&83.17&71.95&67.91&65.22&65.56\\
2.8&77.51&56.32&46.99&---&41.41&\textit{41.86}\\
2.9&55.53&37.77&29.96&27.36&25.29&25.85\\
3.0&39.77&24.99&18.42&---&14.48&\textit{15.02}\\
\end{tabular}
\end{ruledtabular}
\end{table}
\begin{table}
\begin{ruledtabular}
\setlength\extrarowheight{-2pt}
\scriptsize
\caption{
Potential energy $E(r,\theta)$ in meV for the basis sets 
$\zeta=$ 2 through 5 and for the
complete-basis-set extrapolations of
Eqs.~(\ref{eq:234}) and (\ref{eq:345})
 for $\theta=$
75
$^\circ$.
}
\label{table:sup75}
\begin{tabular}
{rrrrrrrr}
$r$&
$E(2\zeta)$&
$E(3\zeta)$&
$E(4\zeta)$&
$E(5\zeta)$&
$E(234)$&
$E(345)$\\
\hline
2.4&255.9&214.7&193.4&184.6&180.4&179.5\\
2.5&184.3&149.4&132.1&125.2&121.6&121.3\\
2.6&133.59&104.02&89.85&---&81.29&\textit{81.33}\\
2.7&97.47&72.53&60.84&56.64&53.79&54.19\\
2.8&71.59&50.65&40.94&---&35.09&\textit{35.62}\\
2.9&52.93&35.42&27.30&24.63&22.40&23.07\\
3.0&39.37&24.77&17.95&---&13.84&\textit{14.49}\\
\end{tabular}
\end{ruledtabular}
\end{table}
\begin{table}
\begin{ruledtabular}
\setlength\extrarowheight{-2pt}
\scriptsize
\caption{
Potential energy $E(r,\theta)$ in meV for the basis sets 
$\zeta=$ 2 through 5 and for the
complete-basis-set extrapolations of
Eqs.~(\ref{eq:234}) and (\ref{eq:345})
 for $\theta=$
80
$^\circ$.
}
\label{table:sup80}
\begin{tabular}
{rrrrrrrr}
$r$&
$E(2\zeta)$&
$E(3\zeta)$&
$E(4\zeta)$&
$E(5\zeta)$&
$E(234)$&
$E(345)$\\
\hline
2.4&244.2&201.7&178.2&168.3&163.9&162.5\\
2.5&178.0&142.2&123.1&115.5&111.5&111.0\\
2.6&131.15&100.92&85.41&---&75.96&\textit{75.90}\\
2.7&97.71&72.32&59.57&54.92&51.82&52.22\\
2.8&73.66&52.39&41.83&---&35.43&\textit{36.01}\\
2.9&56.19&38.39&29.60&26.71&24.28&25.03\\
3.0&43.36&28.45&21.11&---&16.65&\textit{17.40}\\
\end{tabular}
\end{ruledtabular}
\end{table}
\begin{table}
\begin{ruledtabular}
\setlength\extrarowheight{-2pt}
\scriptsize
\caption{
Potential energy $E(r,\theta)$ in meV for the basis sets 
$\zeta=$ 2 through 5 and for the
complete-basis-set extrapolations of
Eqs.~(\ref{eq:234}) and (\ref{eq:345})
 for $\theta=$
85
$^\circ$.
}
\label{table:sup85}
\begin{tabular}
{rrrrrrrr}
$r$&
$E(2\zeta)$&
$E(3\zeta)$&
$E(4\zeta)$&
$E(5\zeta)$&
$E(234)$&
$E(345)$\\
\hline
2.4&262.2&217.2&190.3&178.4&173.7&171.4\\
2.5&192.7&155.0&133.2&124.1&119.7&118.8\\
2.6&143.62&111.93&94.20&---&83.31&\textit{83.00}\\
2.7&108.67&82.10&67.62&62.23&58.75&59.09\\
2.8&83.56&61.29&49.39&---&42.11&\textit{42.68}\\
2.9&65.28&46.59&36.76&33.47&30.75&31.56\\
3.0&51.78&36.04&27.87&---&22.89&\textit{23.72}\\
\end{tabular}
\end{ruledtabular}
\end{table}
\begin{table}
\begin{ruledtabular}
\setlength\extrarowheight{-2pt}
\scriptsize
\caption{
Potential energy $E(r,\theta)$ in meV for the basis sets 
$\zeta=$ 2 through 5 and for the
complete-basis-set extrapolations of
Eqs.~(\ref{eq:234}) and (\ref{eq:345})
 for $\theta=$
90
$^\circ$.
}
\label{table:sup90}
\begin{tabular}
{rrrrrrrr}
$r$&
$E(2\zeta)$&
$E(3\zeta)$&
$E(4\zeta)$&
$E(5\zeta)$&
$E(234)$&
$E(345)$\\
\hline
2.4&312.3&263.5&231.7&216.5&211.8&207.7\\
2.5&230.2&189.4&163.6&152.3&147.6&145.7\\
2.6&172.3&138.2&117.3&---&104.3&\textit{103.4}\\
2.7&131.29&102.71&85.72&79.19&75.22&75.38\\
2.8&101.95&77.96&64.10&---&55.56&\textit{56.06}\\
2.9&80.70&60.48&49.12&45.23&42.13&42.97\\
3.0&65.03&47.89&38.52&---&32.78&\textit{33.68}\\
\end{tabular}
\end{ruledtabular}
\end{table}
\begin{table}
\begin{ruledtabular}
\setlength\extrarowheight{-2pt}
\scriptsize
\caption{
Potential energy $E(r,\theta)$ in meV for the basis sets 
$\zeta=$ 2 through 5 and for the
complete-basis-set extrapolations of
Eqs.~(\ref{eq:234}) and (\ref{eq:345})
 for $\theta=$
95
$^\circ$.
}
\label{table:sup95}
\begin{tabular}
{rrrrrrrr}
$r$&
$E(2\zeta)$&
$E(3\zeta)$&
$E(4\zeta)$&
$E(5\zeta)$&
$E(234)$&
$E(345)$\\
\hline
2.4&401.4&347.1&308.6&288.8&284.5&277.1\\
2.5&295.3&250.2&219.1&204.3&199.7&195.7\\
2.6&220.6&183.0&157.8&---&142.0&\textit{139.9}\\
2.7&167.9&136.4&115.9&107.7&103.2&102.9\\
2.8&130.46&103.93&87.34&---&77.04&\textit{77.32}\\
2.9&103.53&81.08&67.58&62.82&59.23&60.06\\
3.0&83.84&64.70&53.66&---&46.86&\textit{47.82}\\
\end{tabular}
\end{ruledtabular}
\end{table}
\begin{table}
\begin{ruledtabular}
\setlength\extrarowheight{-2pt}
\scriptsize
\caption{
Potential energy $E(r,\theta)$ in meV for the basis sets 
$\zeta=$ 2 through 5 and for the
complete-basis-set extrapolations of
Eqs.~(\ref{eq:234}) and (\ref{eq:345})
 for $\theta=$
100
$^\circ$.
}
\label{table:sup100}
\begin{tabular}
{rrrrrrrr}
$r$&
$E(2\zeta)$&
$E(3\zeta)$&
$E(4\zeta)$&
$E(5\zeta)$&
$E(234)$&
$E(345)$\\
\hline
2.4&541.8&480.1&432.4&405.9&402.3&390.4\\
2.5&396.9&345.7&307.7&287.8&283.7&276.1\\
2.6&294.7&252.1&221.4&---&202.0&\textit{197.5}\\
2.7&222.7&187.1&162.1&151.3&146.5&145.0\\
2.8&171.8&141.8&121.6&---&108.9&\textit{108.6}\\
2.9&135.61&110.13&93.73&87.73&83.52&84.24\\
3.0&109.40&87.60&74.28&---&66.03&\textit{66.99}\\
\end{tabular}
\end{ruledtabular}
\end{table}
\begin{table}
\begin{ruledtabular}
\setlength\extrarowheight{-2pt}
\scriptsize
\caption{
Potential energy $E(r,\theta)$ in meV for the basis sets 
$\zeta=$ 2 through 5 and for the
complete-basis-set extrapolations of
Eqs.~(\ref{eq:234}) and (\ref{eq:345})
 for $\theta=$
105
$^\circ$.
}
\label{table:sup105}
\begin{tabular}
{rrrrrrrr}
$r$&
$E(2\zeta)$&
$E(3\zeta)$&
$E(4\zeta)$&
$E(5\zeta)$&
$E(234)$&
$E(345)$\\
\hline
2.4&753.0&681.1&620.2&585.1&581.5&564.5\\
2.5&548.4&489.2&441.6&414.7&411.4&398.9\\
2.6&403.9&354.8&316.8&---&292.8&\textit{284.6}\\
2.7&302.1&261.1&230.4&215.7&211.0&207.1\\
2.8&230.5&195.9&170.9&---&155.3&\textit{153.4}\\
2.9&179.8&150.4&130.2&122.3&117.5&117.7\\
3.0&143.54&118.33&101.99&---&91.81&\textit{92.53}\\
\end{tabular}
\end{ruledtabular}
\end{table}
\begin{table}
\begin{ruledtabular}
\setlength\extrarowheight{-2pt}
\scriptsize
\caption{
Potential energy $E(r,\theta)$ in meV for the basis sets 
$\zeta=$ 2 through 5 and for the
complete-basis-set extrapolations of
Eqs.~(\ref{eq:234}) and (\ref{eq:345})
 for $\theta=$
110
$^\circ$.
}
\label{table:sup110}
\begin{tabular}
{rrrrrrrr}
$r$&
$E(2\zeta)$&
$E(3\zeta)$&
$E(4\zeta)$&
$E(5\zeta)$&
$E(234)$&
$E(345)$\\
\hline
2.4&1\,063.3&977.8&897.4&851.7&845.9&825.0\\
2.5&770.3&700.3&639.0&603.1&600.0&582.1\\
2.6&562.5&504.6&456.8&---&426.5&\textit{413.4}\\
2.7&416.0&367.7&329.6&309.3&305.5&297.3\\
2.8&312.9&272.3&241.5&---&222.1&\textit{217.4}\\
2.9&240.4&205.9&180.9&170.3&165.2&164.0\\
3.0&189.1&159.5&139.3&---&126.6&\textit{126.5}\\
\end{tabular}
\end{ruledtabular}
\end{table}
\begin{table}
\begin{ruledtabular}
\setlength\extrarowheight{-2pt}
\scriptsize
\caption{
Potential energy $E(r,\theta)$ in meV for the basis sets 
$\zeta=$ 2 through 5 and for the
complete-basis-set extrapolations of
Eqs.~(\ref{eq:234}) and (\ref{eq:345})
 for $\theta=$
120
$^\circ$.
}
\label{table:sup120}
\begin{tabular}
{rrrrrrrr}
$r$&
$E(2\zeta)$&
$E(3\zeta)$&
$E(4\zeta)$&
$E(5\zeta)$&
$E(234)$&
$E(345)$\\
\hline
2.4&2\,158.8&2\,028.8&1\,876.1&1\,803.7&1\,777.3&1\,761.5\\
2.5&1\,551.1&1\,447.3&1\,336.1&1\,277.2&1\,264.5&1\,242.7\\
2.6&1\,116.6&1\,032.3&950.0&---&897.2&\textit{876.9}\\
2.7&808.6&739.1&676.7&639.7&636.9&618.1\\
2.8&591.6&533.6&485.2&---&454.4&\textit{440.6}\\
2.9&439.4&390.6&352.1&331.2&327.7&319.0\\
3.0&332.8&291.4&260.3&---&240.7&\textit{235.6}\\
\end{tabular}
\end{ruledtabular}
\end{table}
\begin{table}
\begin{ruledtabular}
\setlength\extrarowheight{-2pt}
\scriptsize
\caption{
Potential energy $E(r,\theta)$ in meV for the basis sets 
$\zeta=$ 2 through 5 and for the
complete-basis-set extrapolations of
Eqs.~(\ref{eq:234}) and (\ref{eq:345})
 for $\theta=$
130
$^\circ$.
}
\label{table:sup130}
\begin{tabular}
{rrrrrrrr}
$r$&
$E(2\zeta)$&
$E(3\zeta)$&
$E(4\zeta)$&
$E(5\zeta)$&
$E(234)$&
$E(345)$\\
\hline
2.4&4\,345.8&4\,126.9&3\,822.2&3\,711.0&3\,623.7&3\,646.3\\
2.5&3\,111.7&2\,945.0&2\,728.0&2\,638.1&2\,587.0&2\,585.7\\
2.6&2\,221.8&2\,091.3&1\,936.2&---&1\,835.8&\textit{1\,824.3}\\
2.7&1\,586.4&1\,482.0&1\,369.8&1\,310.6&1\,297.6&1\,275.9\\
2.8&1\,136.8&1\,051.5&968.9&---&916.0&\textit{895.6}\\
2.9&820.8&750.2&687.8&650.7&648.0&629.0\\
3.0&599.9&540.9&492.5&---&461.7&\textit{447.7}\\
\end{tabular}
\end{ruledtabular}
\end{table}
\begin{table}
\begin{ruledtabular}
\setlength\extrarowheight{-2pt}
\scriptsize
\caption{
Potential energy $E(r,\theta)$ in meV for the basis sets 
$\zeta=$ 2 through 5 and for the
complete-basis-set extrapolations of
Eqs.~(\ref{eq:234}) and (\ref{eq:345})
 for $\theta=$
140
$^\circ$.
}
\label{table:sup140}
\begin{tabular}
{rrrrrrrr}
$r$&
$E(2\zeta)$&
$E(3\zeta)$&
$E(4\zeta)$&
$E(5\zeta)$&
$E(234)$&
$E(345)$\\
\hline
2.4&8\,390.1&7\,978.5&7\,389.7&7\,212.6&7\,005.7&7\,110.0\\
2.5&6\,009.8&5\,718.9&5\,298.9&5\,160.7&5\,024.9&5\,080.5\\
2.6&4\,279.8&4\,065.7&3\,769.2&---&3\,576.0&\textit{3\,601.9}\\
2.7&3\,035.9&2\,872.9&2\,663.9&2\,575.7&2\,528.1&2\,524.2\\
2.8&2\,150.5&2\,022.7&1\,874.3&---&1\,778.4&\textit{1\,765.3}\\
2.9&1\,525.7&1\,423.1&1\,316.2&1\,258.7&1\,247.4&1\,225.0\\
3.0&1\,088.1&1\,004.2&925.4&---&874.9&\textit{854.6}\\
\end{tabular}
\end{ruledtabular}
\end{table}
\begin{table}
\begin{ruledtabular}
\setlength\extrarowheight{-2pt}
\scriptsize
\caption{
Potential energy $E(r,\theta)$ in meV for the basis sets 
$\zeta=$ 2 through 5 and for the
complete-basis-set extrapolations of
Eqs.~(\ref{eq:234}) and (\ref{eq:345})
 for $\theta=$
150
$^\circ$.
}
\label{table:sup150}
\begin{tabular}
{rrrrrrrr}
$r$&
$E(2\zeta)$&
$E(3\zeta)$&
$E(4\zeta)$&
$E(5\zeta)$&
$E(234)$&
$E(345)$\\
\hline
2.4&14\,995.&14\,146.&13\,127.&12\,834.&12\,468.&12\,664.\\
2.5&10\,759.4&10\,212.2&9\,461.9&9\,244.3&8\,973.2&9\,118.2\\
2.6&7\,666.2&7\,294.8&6\,760.4&---&6\,411.6&\textit{6\,504.2}\\
2.7&5\,430.1&5\,165.1&4\,790.8&4\,661.8&4\,546.6&4\,586.8\\
2.8&3\,829.6&3\,632.7&3\,371.8&---&3\,202.0&\textit{3\,216.6}\\
2.9&2\,695.0&2\,543.7&2\,361.0&2\,279.3&2\,242.6&2\,231.6\\
3.0&1\,897.3&1\,777.9&1\,648.3&---&1\,564.8&\textit{1\,547.8}\\
\end{tabular}
\end{ruledtabular}
\end{table}
\begin{table}
\begin{ruledtabular}
\setlength\extrarowheight{-2pt}
\scriptsize
\caption{
Potential energy $E(r,\theta)$ in meV for the basis sets 
$\zeta=$ 2 through 5 and for the
complete-basis-set extrapolations of
Eqs.~(\ref{eq:234}) and (\ref{eq:345})
 for $\theta=$
160
$^\circ$.
}
\label{table:sup160}
\begin{tabular}
{rrrrrrrr}
$r$&
$E(2\zeta)$&
$E(3\zeta)$&
$E(4\zeta)$&
$E(5\zeta)$&
$E(234)$&
$E(345)$\\
\hline
2.4&23\,864.&22\,161.&20\,696.&20\,203.&19\,764.&19\,917.\\
2.5&17\,130.&16\,100.&14\,945.&14\,604.&14\,199.&14\,406.\\
2.6&12\,218.&11\,570.&10\,719.&---&10\,166.&\textit{10\,321.}\\
2.7&8\,658.0&8\,225.8&7\,624.3&7\,441.1&7\,232.3&7\,334.9\\
2.8&6\,100.7&5\,796.6&5\,379.3&---&5\,107.5&\textit{5\,165.8}\\
2.9&4\,281.1&4\,057.7&3\,769.6&3\,660.2&3\,582.5&3\,596.5\\
3.0&2\,997.7&2\,828.0&2\,628.2&---&2\,498.8&\textit{2\,495.9}\\
\end{tabular}
\end{ruledtabular}
\end{table}
\begin{table}
\begin{ruledtabular}
\setlength\extrarowheight{-2pt}
\scriptsize
\caption{
Potential energy $E(r,\theta)$ in meV for the basis sets 
$\zeta=$ 2 through 5 and for the
complete-basis-set extrapolations of
Eqs.~(\ref{eq:234}) and (\ref{eq:345})
 for $\theta=$
170
$^\circ$.
}
\label{table:sup170}
\begin{tabular}
{rrrrrrrr}
$r$&
$E(2\zeta)$&
$E(3\zeta)$&
$E(4\zeta)$&
$E(5\zeta)$&
$E(234)$&
$E(345)$\\
\hline
2.4&32\,425.&29\,636.&27\,910.&27\,167.&26\,839.&26\,734.\\
2.5&23\,243.&21\,575.&20\,104.&19\,613.&19\,166.&19\,328.\\
2.6&16\,584.&15\,573.&14\,441.&---&13\,710.&\textit{13\,869.}\\
2.7&11\,755.6&11\,114.2&10\,297.0&10\,059.1&9\,766.3&9\,921.4\\
2.8&8\,282.2&7\,850.3&7\,280.8&---&6\,910.5&\textit{7\,008.9}\\
2.9&5\,806.2&5\,500.2&5\,109.0&4\,974.6&4\,854.9&4\,896.5\\
3.0&4\,056.8&3\,831.6&3\,562.9&---&3\,388.9&\textit{3\,403.6}\\
\end{tabular}
\end{ruledtabular}
\end{table}
\begin{table}
\begin{ruledtabular}
\setlength\extrarowheight{-2pt}
\scriptsize
\caption{
Potential energy $E(r,\theta)$ in meV for the basis sets 
$\zeta=$ 2 through 5 and for the
complete-basis-set extrapolations of
Eqs.~(\ref{eq:234}) and (\ref{eq:345})
 for $\theta=$
180
$^\circ$.
}
\label{table:sup180}
\begin{tabular}
{rrrrrrrr}
$r$&
$E(2\zeta)$&
$E(3\zeta)$&
$E(4\zeta)$&
$E(5\zeta)$&
$E(234)$&
$E(345)$\\
\hline
2.4&36\,142.&32\,814.&31\,025.&30\,152.&29\,929.&29\,643.\\
2.5&25\,879.&23\,879.&22\,298.&21\,730.&21\,298.&21\,400.\\
2.6&18\,462.&17\,262.&16\,018.&---&15\,218.&\textit{15\,356.}\\
2.7&13\,088.&12\,338.&11\,430.&11\,167.&10\,842.&11\,015.\\
2.8&9\,219.7&8\,722.9&8\,088.0&---&7\,675.6&\textit{7\,789.4}\\
2.9&6\,461.2&6\,114.4&5\,678.3&5\,533.8&5\,395.3&5\,449.8\\
3.0&4\,511.4&4\,259.4&3\,960.8&---&3\,767.5&\textit{3\,790.8}\\
\end{tabular}
\end{ruledtabular}
\end{table}

\begin{table}
\begin{ruledtabular}
\setlength\extrarowheight{-2pt}
\scriptsize
\caption{
Positions in \AA~of the atoms for  
$S=13$.
}
\label{table:S13positions}
\begin{tabular}
{cccc}
&
$x$&
$y$&
$z$\\
\hline
F&2.11&2.11&0.00\\
Ba&0.59&0.59&0.00\\
Ne&-4.53&0.00&0.00\\
Ne&0.00&-4.53&0.00\\
Ne&-0.01&-0.01&-4.65\\
Ne&-0.01&-0.01&4.65\\
Ne&0.00&4.61&0.00\\
Ne&4.61&0.00&0.00\\
Ne&-4.44&-2.20&-2.22\\
Ne&-4.44&-2.20&2.22\\
Ne&-2.20&-4.44&-2.22\\
Ne&-2.20&-4.44&2.22\\
Ne&-2.21&-2.21&-4.46\\
Ne&-2.21&-2.21&4.46\\
Ne&-4.45&2.22&-2.22\\
Ne&-4.45&2.22&2.22\\
Ne&2.22&-4.45&-2.22\\
Ne&2.22&-4.45&2.22\\
Ne&-2.22&2.22&-4.47\\
Ne&-2.22&2.22&4.47\\
Ne&2.22&-2.22&-4.47\\
Ne&2.22&-2.22&4.47\\
Ne&-2.21&4.46&-2.22\\
Ne&-2.21&4.46&2.22\\
Ne&4.46&-2.21&-2.22\\
Ne&4.46&-2.21&2.22\\
Ne&2.21&2.21&-4.43\\
Ne&2.21&2.21&4.43\\
Ne&2.21&4.45&-2.21\\
Ne&2.21&4.45&2.21\\
Ne&4.45&2.21&-2.21\\
Ne&4.45&2.21&2.21\\
Ne&-4.44&-4.44&0.00\\
Ne&-4.45&0.00&-4.45\\
Ne&-4.45&0.00&4.45\\
Ne&0.00&-4.45&-4.45\\
Ne&0.00&-4.45&4.45\\
Ne&-4.45&4.45&0.00\\
Ne&4.45&-4.45&0.00\\
Ne&0.00&4.45&-4.45\\
Ne&0.00&4.45&4.45\\
Ne&4.45&0.00&-4.45\\
Ne&4.45&0.00&4.45\\
Ne&4.41&4.41&0.00\\
Ne&-6.71&-2.23&0.00\\
Ne&-2.23&-6.71&0.00\\
Ne&-6.71&0.00&-2.24\\
Ne&-6.71&0.00&2.24\\
Ne&0.00&-6.71&-2.24\\
Ne&0.00&-6.71&2.24\\
Ne&-6.71&2.24&0.00\\
Ne&2.24&-6.71&0.00\\
Ne&-2.26&0.00&-6.75\\
Ne&-2.26&0.00&6.75\\
Ne&0.00&-2.26&-6.75\\
Ne&0.00&-2.26&6.75\\
Ne&0.00&2.26&-6.74\\
Ne&0.00&2.26&6.74\\
Ne&2.26&0.00&-6.74\\
Ne&2.26&0.00&6.74\\
Ne&-2.25&6.73&0.00\\
Ne&6.73&-2.25&0.00\\
Ne&0.00&6.73&-2.25\\
Ne&0.00&6.73&2.25\\
Ne&6.73&0.00&-2.25\\
Ne&6.73&0.00&2.25\\
Ne&2.25&6.71&0.00\\
Ne&6.71&2.25&0.00\\
Ne&-4.45&-4.45&-4.45\\
Ne&-4.45&-4.45&4.45\\
Ne&-4.46&4.46&-4.46\\
Ne&-4.46&4.46&4.46\\
Ne&4.46&-4.46&-4.46\\
Ne&4.46&-4.46&4.46\\
Ne&4.45&4.45&-4.45\\
Ne&4.45&4.45&4.45
\end{tabular}
\end{ruledtabular}
\end{table}

\bibliography{BaFAr.bib}

\end{document}